\documentclass[journal]{IEEEtran}
\IEEEoverridecommandlockouts
\usepackage[labelsep=period]{caption}
\usepackage{subfig}
\usepackage{cite}
\usepackage{amsmath,amssymb,amsfonts}
\usepackage{algorithmic}
\usepackage{graphicx}
\usepackage{textcomp}
\usepackage{xcolor}
\usepackage[ruled,linesnumbered]{algorithm2e}
\usepackage{amsthm}
\usepackage{bm}
\usepackage{comment}
\usepackage{array}
\usepackage{booktabs}
\usepackage{enumitem}
\def\BibTeX{{\rm B\kern-.05em{\sc i\kern-.025em b}\kern-.08em
    T\kern-.1667em\lower.7ex\hbox{E}\kern-.125emX}}
\makeatletter
\renewenvironment{proof}[1][\proofname]{\par
  \pushQED{\qed}%
  \normalfont \topsep6\p@\@plus6\p@\relax
  \trivlist
  \item[\hskip\labelsep
        \itshape #1\@addpunct{:}]\ignorespaces
}{%
  \popQED\endtrivlist\@endpefalse
}
\makeatother

\begin{document}
\title{Drift-Adaptive Slicing-Based Resource Management for Cooperative ISAC Networks}

\author{Shisheng Hu, Jie Gao, Xue Qin, Conghao Zhou, Xinyu Huang, Mushu Li, Mingcheng He, \\and Xuemin (Sherman) Shen, \IEEEmembership{Fellow, IEEE}
\thanks{Shisheng Hu, Xue Qin, Conghao Zhou, Xinyu Huang, Mingcheng He, and Xuemin (Sherman) Shen are with
the Department of Electrical and Computer Engineering, University of
Waterloo, Waterloo, ON N2L 3G1, Canada (email: \{s97hu, x7qin, c89zhou, x357huan, m64he, sshen\}@uwaterloo.ca).}
\thanks{Jie Gao is with the School of Information Technology, Carleton University,
Ottawa, ON K1S 5B6, Canada (email: jie.gao6@carleton.ca).}
\thanks{Mushu Li is with the Department of Computer Science and Engineering, Lehigh University, Bethlehem, PA 18015, USA (email: mul224@lehigh.edu).}
\thanks{Part of this research work was presented at the IEEE/CIC International Conference on Communications in China, 2024~\cite{icccpaper2024}.}
\thanks{\textit{(Corresponding authors: Conghao Zhou, Xue Qin.)}}
}
\maketitle
\begin{abstract}
In this paper, we propose a {novel drift-adaptive slicing-based resource management scheme} for cooperative integrated sensing and communication (ISAC) networks. Particularly, we establish two {network slices} to provide sensing and communication services, respectively. In the large-timescale planning for the slices, we partition the sensing region of interest (RoI) of each mobile device and reserve network resources accordingly, {facilitating low-complexity distance-based sensing target assignment in small timescales.} To cope with the non-stationary spatial distributions of mobile devices and sensing targets, which can result in the drift in modeling the distributions and ineffective planning decisions, we construct digital twins (DTs) of the slices. In each DT, a drift-adaptive statistical model and an emulation function are developed for the spatial distributions in the corresponding slice, which facilitates closed-form decision-making and efficient validation of a planning decision, respectively. Numerical results show that the {proposed drift-adaptive slicing-based resource management scheme} can increase the service satisfaction ratio by up to 18\% and reduce resource consumption by up to 13.1\% when compared with benchmark schemes.
\end{abstract}


\section{{Introduction}}

Integrated sensing and communication (ISAC) is recognized as a transformative paradigm in the forthcoming 6G era~\cite{ISAC_survey,10608156}. However, the substantial resource demand and strict performance requirements of ISAC pose a challenge to its implementation. 
Cooperative ISAC\cite{Cooperative_ISAC_Jin_Shi_WC, CooperativeISAC_Christos}, also known as networked ISAC~\cite{Network_ISAC_magazine, Networked_ISAC_xiaodan}, is promising to tackle this challenge. {In cooperative ISAC networks, multiple devices, each equipped with both communication and sensing functions, can be used for communication and sensing service provision. Particularly, multiple devices can be selected to monitor a sensing target, exploiting their respective strengths, {such as the proximity of mobile devices to the target or the abundant resources available at access points (APs).} 
To handle the potential high complexity of multi-device and multi-service coordination in cooperative ISAC networks, slicing-based resource management, particularly, the two-stage slicing scheme~\cite{wen_wu_slicing,ConghaoDT}, can be exploited. Over large timescales, referred to as planning windows, network slices for the individual services can be proactively configured through network resource reservation, a process referred to as network planning~\cite{ConghaoDT}, which reduces the complexity of small-timescale resource management for individual slices.} 

{Despite the potential, the challenge of leveraging the slicing-based resource management scheme for cooperative ISAC networks is to efficiently and accurately characterize the random spatial distributions of mobile devices and targets.} A statistical spatial model, such as a spatial point process, is commonly adopted due to its capability to derive closed-form estimation of the expected service demands and capacities for network planning~\cite{SG_ISAC_time_allocation, SG_OJCS_jounal_ISAC, sun2024performance, meng2023network}.
However, in temporally non-stationary environments, across planning windows, the parameters in a statistical spatial model that best fit the actual spatial distributions would change. If the change is drastic, the parameter prediction for the subsequent planning window may suffer from severe accuracy degradation~\cite{model_drift_wireless}, which causes the \textit{drift} of the spatial model. {Especially, a spatial model with finer granularity, defined as the level of detail or spatial resolution at which spatial distribution information is captured, would generally face a higher risk of the drift because the model is typically characterized by more parameters.} For example, a cluster point process can be used to characterize the spatial correlation between mobile devices and targets~\cite{SG_ISAC_time_allocation}, which makes it more accurate in most cases but less robust to a sudden change of the spatial correlation, compared with the use of two independent point processes~\cite{SG_OJCS_jounal_ISAC, sun2024performance, meng2023network}. 
Digital twins (DTs) have received increasing attention for modeling and managing wireless networks. DTs, as digital representations of individual network entities or network slices, can help facilitate configurable collection and processing of their data and efficient and accurate modeling of them~\cite{shen2021holistic}. Particularly, introducing DTs of communication and sensing slices for spatial modeling in cooperative ISAC networks has the following advantages to tackle the drift issue:
\begin{itemize}
\item \textit{Efficient Data Utilization:} Through sensing service provision, the mobile devices and APs continuously obtain sensing data of sensing targets. Such sensing data, along with the location data of mobile devices, can be stored in the slice DTs and used for spatial modeling.
\item \textit{Statistical Model Adaptation:} The collected data in the DTs can be used to continuously optimize the statistical modeling of the spatial distributions, such as the model granularity determination and the model parameter prediction.
\item \textit{Emulation-Based Model Validation:} Due to temporal correlation, historical snapshots of the spatial distributions can be used to emulate the distributions in the following planning window. Such emulation can be used to validate whether a statistical spatial model has adapted to the drift that occurred earlier. 
\end{itemize}


In this paper, we propose {a drift-adaptive slicing-based resource management scheme for cooperative ISAC networks}. 
Our objective is to efficiently realize the collaboration of APs and mobile devices in the communication and sensing service provision, such that the minimum spectrum and edge computing resource is consumed in meeting the service demands. To achieve this objective, we propose a slicing-based resource management scheme, where separate network slices are created and managed to provide the communication and sensing services, respectively. In the large-timescale planning of the slices, the sensing region of interest (RoI) of each mobile device is partitioned for low-complexity distance-based sensing target assignment in small timescales. Then, we formulate an optimization problem that minimizes the resource consumption in service provision by properly determining the sensing RoI partitioning and the resource reservation for the slices. By collecting and processing the historical spatial distributions of mobile devices and targets, we construct DTs of the network slices to obtain drift-adaptive statistical models and emulation functions for their spatial distributions in individual slices. Leveraging the statistical spatial models, we derive closed-form formulas in solving the optimization problem. In addition, via emulation, the reliability of a statistical spatial model and the corresponding network planning decision are validated. 
The main contributions of this paper are summarized as follows. 

\begin{itemize}
\item {We propose a slicing-based resource management scheme} that enables low-complexity coordination of mobile devices and APs within ISAC networks by large-timescale network planning including sensing RoI partitioning and resource reservation.
\item We construct DTs of network slices that provide drift-adaptive statistical models and emulation functions for characterizing the spatial distributions of sensing targets and mobile devices. Based on the DTs, we derive a closed-form and reliable network planning solution.
\end{itemize} 

The remainder of this paper is organized as follows. In Section~\ref{realted_work}, we review the related works. In Section~\ref{sec_system_model}, we present the considered scenario, {our proposed slicing-based resource management scheme for cooperative ISAC networks}, and the network planning problem formulation. The establishment of DTs of network slices is introduced in Section~\ref{chap_sliceDT}. The network planning algorithm based on the slice DTs is introduced in Section~\ref{overall_scheme_section}. Simulation results are provided in Section~\ref{simulation_sec}, followed by the conclusion of this paper in Section~\ref{sec_conclusion}.

\section{Related Works}\label{realted_work}

Spatial information of mobile devices and users plays a critical role in managing emerging wireless networks and systems such as 
industrial IoT networks~\cite{10757329} and mobile edge computing systems~\cite{Mobility_MEC, Edge_cache_Lyu}. Particularly, information on the spatial distributions of devices and targets is crucial for {resource management for realizing the ISAC paradigm.}
The real-time spatial distributions empower small-timescale resource management for ISAC, e.g., power control~\cite{unified_ISAC_resource_allocation,JSAC_joint_beamforming_power}, and spectrum allocation~\cite{Jie_chen_dual}. For large-timescale network planning, spatial point processes that are temporally stationary, i.e., those with constant parameters over time, are commonly used to characterize the spatial distributions~\cite{new_trend_SG_survey}. Accordingly, the formulas for estimating spatially averaged sensing and communication performances, e.g., target detection probability and ergodic transmission rate, can be derived for performance analysis and network planning~\cite{Letter_stochastic_radar_network, TVT_stochastic_interference}.

Spatial information is even more critical for{ the resource management in cooperative ISAC networks.} In cooperative ISAC networks, multiple devices can be used for monitoring individual sensing targets. The information of their spatial distributions, along with that of the sensing targets, determine the optimal coordination of the devices and the corresponding resource allocation. With the real-time spatial distribution information, the selection and beamforming designs for multiple APs was investigated in~\cite{Network_ISAC_magazine, sensing_node_selection_ISAC}, and the fusion scheme of the sensing data from the multiple APs was investigated in~\cite{Cooperative_ISAC_Jin_Shi_WC}. For large-timescale network planning, the spatial distributions are mainly modeled by temporally stationary spatial point processes. Based on such a spatial model, the density of APs~\cite{SG_OJCS_jounal_ISAC, sun2024performance} and the clustering of APs for interference management~\cite{meng2023network} in multi-cell dense wireless networks can be efficiently optimized. In addition, time allocation was optimized in time-division-based ISAC systems, where APs and mobile devices cooperate in sensing service provision in a bi-static sensing mode~\cite{SG_ISAC_time_allocation}. Different from the works, for reduced coordination complexity, we investigate the large-timescale network planning for the cooperation of mobile devices and APs in mono-static sensing mode, establishing network slice DTs for the network planning in temporally non-stationary environments.

{DTs are envisioned as a promising paradigm to enhance the service provision in wireless communication networks~\cite{shen2021holistic}. Existing works were mainly focused on establishing (i) user-level DTs to characterize the real-time user status and service demand (e.g., task arrival pattern in mobile edge computing applications~\cite{Shisheng_DNNinference_DT} and sensing devices' data quality in collaborative sensing scenarios~\cite{sensing_mushu}), or (ii) network-level DTs to represent the real-time network status using proper data structures such as graphs, and to evaluate service provision using network simulators~\cite{Network_DT_Comm_Mag, DT_slice_based_comm_mag} or learning-based algorithms~\cite{DT_slice, Network_DT_Comm_Mag, DT_slice_based_comm_mag}. In this paper, we focus on developing network-level DTs, particularly the DTs of network slices for communication and sensing services in a cooperative ISAC network, for efficient and reliable large-timescale planning of the network slices. To enhance the efficiency, instead of aggregating the real-time status of individual network entities, we investigate their statistical modeling, which facilitates the derivation of closed-form formulas for large-timescale planning. To enhance the reliability, instead of relying on a single statistical model, which is vulnerable to model drift, we investigate the ensemble of two statistical models and utilize network emulation to evaluate network planning decisions under potential model drift before their actual implementation.}

\section{System Model and Problem Formulation}\label{sec_system_model}


\begin{figure}
    \centering
    \includegraphics[width=8.5cm]{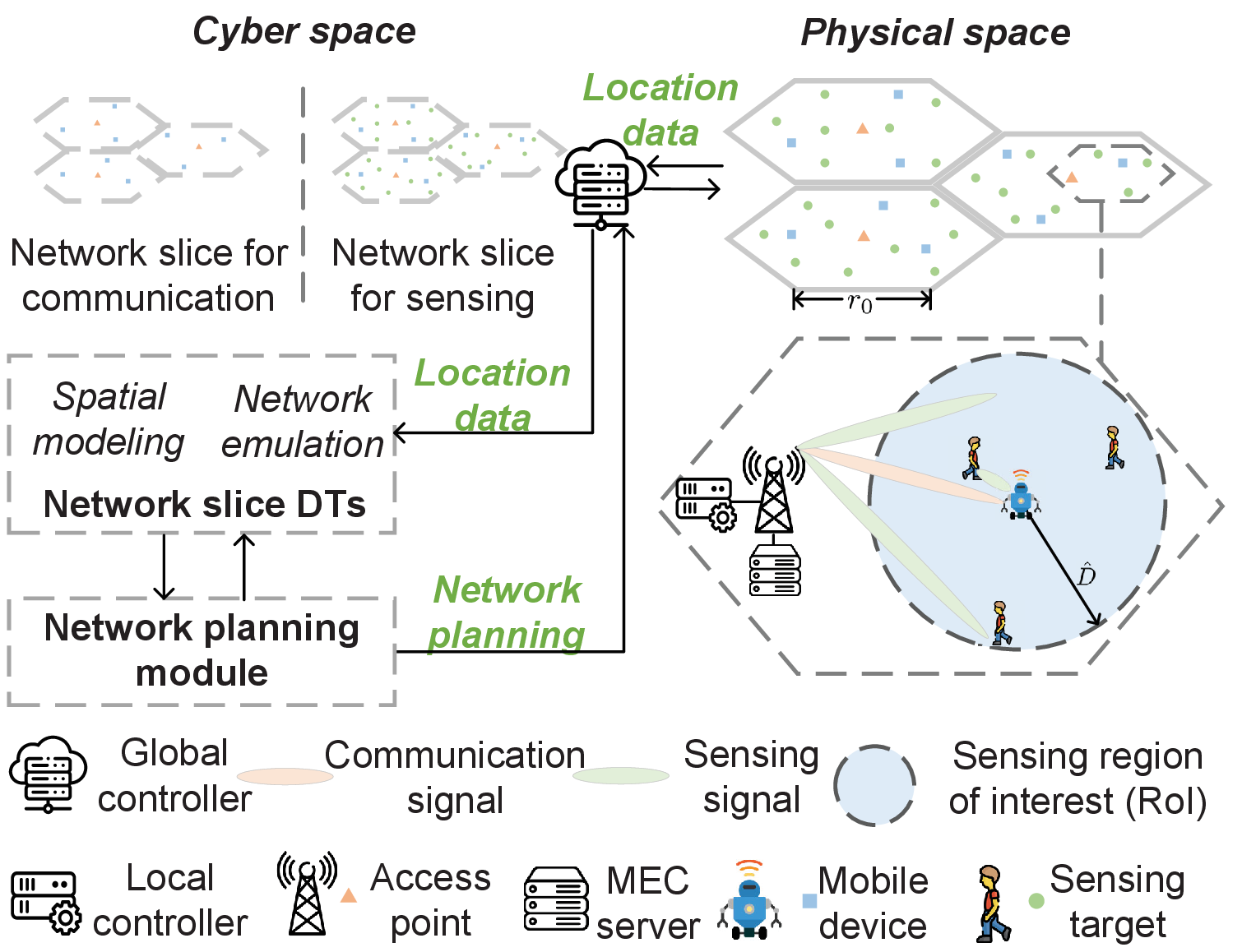}
    \caption{{The considered scenario.}}
    \label{system_overview}
\end{figure}



\subsection{Considered Scenario}
As shown in Fig.~\ref{system_overview}, we consider an industrial zone, which is segmented to $L$ non-overlapping service provision areas. Each area is a hexagon with a side length $r_0$ and a size of $A_0 = 3\sqrt{3}r_0^2/2$. In each area, there exist sensing targets, mobile devices, and an AP. The service demands and capabilities of them are introduced as follows.
\begin{itemize}
\item \textbf{Mobile Devices}: such as load-carrying and inspection robots, that have sensing and communication demands and capabilities within the ISAC paradigm~\cite{wang2024adaptive}. Each mobile device monitors the targets in their sensing RoI which is a circle centered at it with radius $\hat{D}$. In addition, the mobile devices have on-device computing resources that can be used for sensing data processing. 
\item \textbf{Sensing Targets}: such as human workers, that need to be detected to enhance workspace safety and efficiency. {In addition, their state, such as the location, velocity, and activity, needs to be periodically tracked.} 
\item \textbf{AP}: located in the center of the area, that has sensing and communication capabilities and has access to edge computing resources that can be used for sensing data processing. 
\end{itemize}

{We consider a cooperative ISAC network, where mobile devices and APs collaborate in sensing and communication service provision. In this network, the respective advantages of mobile devices and APs in sensing, such as closer proximity of mobile devices to sensing targets and more abundant resources available at APs, are leveraged for effective target assignment. In addition, the scheduling of mobile devices to perform either communication or sensing is optimized under a time-division-based ISAC paradigm, which ensures that the mobile devices can complete the sensing tasks for the assigned targets while completing their data transmission tasks~\cite{ISAC_scheduling}. With such cooperation, the network resources can be efficiently leveraged in the service provision.} 

\subsection{Overview of the Slicing Scheme for Cooperative ISAC}
{We propose the following two-stage slicing-based resource management scheme to efficiently realize the cooperation of mobile device and APs in the cooperative ISAC network.}

\subsubsection{RoI Partitioning-Based Target Assignment}
\begin{figure}
    \centering
    \subfloat[RoI partitioning-based sensing target assignment.]{\includegraphics[width=8cm]{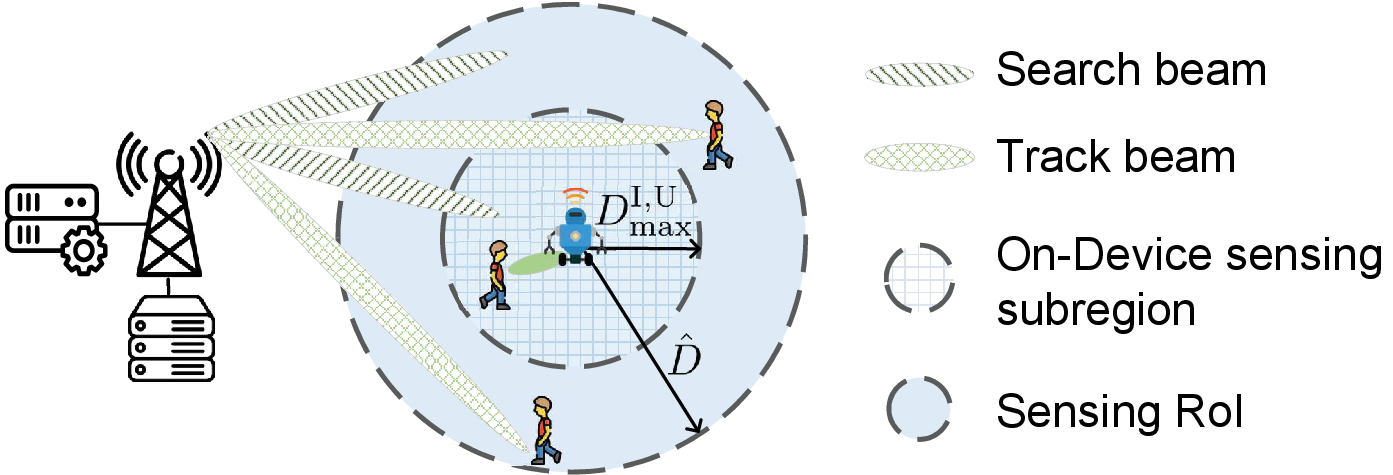}
    \label{Sensing_model}}
    \\
    \subfloat[Two-stage slicing-based resource management scheme.]{\includegraphics[width=7cm]{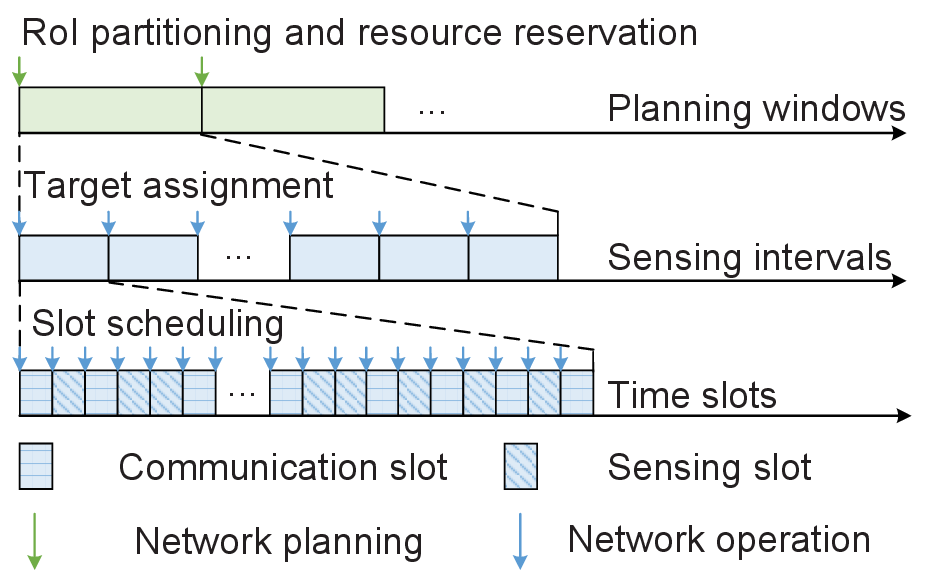}
    \label{time_illust}}
    \caption{The proposed slicing-based resource management scheme for cooperative ISAC networks.}
    \label{fig_combined_roi_time}
\end{figure}

Each AP collaborates with each mobile device in its coverage for monitoring the targets in the sensing RoI of the mobile device. First, the AP periodically detects targets that newly arrive or become active within the RoI of each mobile device. Then, a detected target is assigned to either an AP or a mobile device for tracking according to an RoI partitioning-based rule. Specifically, as shown in Fig.~\ref{Sensing_model}, we draw a concentric inner circle with radius $D^{\rm I,U}_{\max}$ in the RoI of each mobile device and define it as the \textit{on-device sensing subregion}. Due to the proximity, we assign the targets in the on-device sensing subregions to mobile devices and allocate shared spectrum resources for them to track the targets. {Specifically, a target will be assigned to the mobile device if it is {(i)} in the on-device sensing subregion of the mobile device and {(ii)} closest in distance to the mobile device than to any other mobile devices.}
\subsubsection{Two-Stage Slicing-Based Resource Management}
{To efficiently provide the communication and sensing services, the two-stage slicing-based resource management scheme~\cite{ConghaoDT} is adopted.} Particularly, two network slices are established by a global controller to provide sensing and communication services, respectively. As shown in Fig.~\ref{time_illust}, the management of the network slices involves small-timescale network operation by local controllers at the APs and large-timescale network planning by the global controller:
\begin{itemize}
\item \textit{Small-Timescale Network Operation:} At the beginning of each sensing interval that consists of $T$ time slots, the above-mentioned RoI partitioning-based target assignment is carried out. At the beginning of each time slot that lasts $\tau$ seconds, each mobile device in the coverage of an AP is scheduled by the local controller at the AP to perform either mono-static sensing or communication according to pre-determined probabilities{\footnote{The scheduling can also incorporate the instantaneous channel conditions. For example, according to~\cite{ISAC_scheduling,threshold_property_paper}, the optimal communication scheduling based on channel conditions has a threshold property. Given the active probability and the statistical property of the channel conditions, {the threshold can be adjusted to ensure that the proportion of slots when the device is active for communication is approximately equal to the active probability.}}}, to be detailed in Section~\ref{sensing_mode}. 
\item \textit{Large-Timescale Network Planning:} At the beginning of each planning window that consists of $M$ sensing intervals, {the global controller makes a network planning decision including resource reservation and RoI partitioning.} Specifically, spectrum and edge computing resources are reserved, and the probabilities for a mobile device to perform communication and sensing in each time slot are determined. In addition, the radius of the on-device sensing subregions is determined. 
\end{itemize}

{In the proposed two-stage resource management scheme, large-timescale network planning empowers low-complexity real-time sensing target assignment and resource allocation for individual devices, while small-timescale network operation incorporates real-time network dynamics, such as instantaneous channel conditions, to further improve service provision performance and resource utilization efficiency. In this paper, we focus on large-timescale network planning, taking into account the varying spatial distributions of mobile devices and targets across different planning windows.} As shown in Fig.~\ref{system_overview}, the global controller aggregates historical spatial distributions of mobile devices and targets in each network slice to establish DTs of the network slices. The DTs are used to achieve adaptive spatial modeling and network emulation, to be detailed in Section~\ref{chap_sliceDT}. {Based on the DTs, {a network planning algorithm is developed to efficiently and reliably make the network planning decision}, which will be detailed in Section~\ref{overall_scheme_section}.}

\subsection{Communication Model}\label{comm_model}
The global controller reserves $X^{\rm A}_{\rm c}\in \mathbb{Z}_{\geq 0}$ subcarriers for the communication of the mobile devices in the coverage of each AP, where $X^{\rm A}_{\rm c}$ is an optimization variable. The total amount of spectrum resources reserved for the communication in the coverage of the $L$ APs is thus
$Z_{\rm c}^{\rm A} = L\cdot X^{\rm A}_{\rm c}\cdot B_{\rm c,0}$, where $B_{\rm c,0}$ is the bandwidth of each subcarrier in Hertz. We consider the scenario where there exists at least one mobile device in the coverage of each AP. In addition, we consider homogeneity among different mobile devices and APs\footnote{We consider homogeneous mobile devices in industrial scenarios, all with identical capabilities and requirements for executing the same type of tasks. In the scenarios with heterogeneous mobile devices, we can first cluster the mobile devices with similar capabilities and requirements. Then, mobile devices in each cluster can be approximated as homogeneous, and the designs in this paper can be applied.}. We focus on one representative mobile device and one representative AP when deriving the service demands and capacities, and refer to them as the mobile device and the AP, respectively.
The number of mobile devices in the coverage of the AP is denoted by $N^{\rm I}$. At the beginning of each time slot, with an active probability $\rho_{\rm c}\in(0,1]$, which is an optimization variable, each mobile device in the coverage of each AP is scheduled by the local controller at the AP for communication. The number of all mobile devices that are scheduled for communication in the time slot is denoted by $N^{\rm I}_{\rm c}$. Conditioning on that $N^{\rm I}=n$ and the mobile device is scheduled for communication, the probability mass function (PMF) of $N^{\rm I}_{\rm c}$ is
\begin{equation}
\mathbb{P}\{N^{\rm I}_{\rm c} = n_{\rm c}|N^{\rm I}=n\} = \binom{n_{\rm c}-1}{n-1} \rho_{\rm c}^{n_{\rm c}-1} (1 - \rho_{\rm c})^{n - n_{\rm c}},
\end{equation}
where $n_{\rm c} = 1,2,3,...,n$, and $n=1,2,...$. As a result, we have $\mathbb{E}\left[N^{\rm I}_{\rm c}\right|N^{\rm I}]=1+(N^{\rm I}-1)\rho_{\rm c}$.
The subcarriers are evenly allocated to the mobile devices scheduled for communication, and the number of the subcarriers allocated to the mobile device is $X_{\rm c}^{\rm I} = {X^{\rm A}_{\rm c}/}{N^{\rm I}_{\rm c}}$. 

\noindent\textbf{Proposition 1:} \textit{Denote the transmission rate over each subcarrier (in bits per time slot) by $R_0$ and the transmission of the mobile device by $R$. A lower bound of the expectation of transmission rate $R$ is}
\begin{equation}\label{low_bound_R_secIII}
\bar{R} = \frac{R_0X^{\rm A}_{\rm c}}{\mathbb{E}\left[N^{\rm I}\right]+1/\rho_{\rm c}-1}.
\end{equation}
\begin{proof} Please refer to Appendix A.
\end{proof}
The lower bound in~\eqref{low_bound_R_secIII}, i.e., $\bar{R}$, corresponds to the communication service capacity. The average amount of data that each mobile device needs to upload per time slot, denoted by $\hat{R}$, corresponds to the communication service demand. {With a lower active probability for communication $\rho_{\rm c}$, i.e., a higher active probability for sensing, the lower bound of the transmission rate $\bar{R}$ decreases. The reason is that in such cases, the utilization efficiency of the reserved spectrum resources for communication would be reduced. This reflects a potential trade-off between the capacity of the communication service and that of the sensing service, which should be optimized in the network planning.}

\subsection{Sensing Model}\label{sensing_mode}
We consider two kinds of sensing tasks. The first kind is the \textit{target detection task}. Particularly, in each sensing interval, the AP identifies the targets that have newly arrived or become active within the RoIs of the mobile devices in its coverage. At the end of each sensing interval, based on the detection results, the local controller at the AP updates a database that catalogs the targets within the RoIs of the mobile devices in its coverage. The second kind is the \textit{target tracking task}. In particular, within a sensing interval, a target from the database is assigned to either a mobile device or the AP for monitoring across multiple time slots. At the end of each of the time slots, a sensing data sample is collected for the target. At the end of the interval, the accumulated sensing data samples for each target are processed to determine the state of the target. The tracking results are recorded in the target database, providing prior information to facilitate target tracking in subsequent sensing intervals.
\subsubsection{{Target Detection and Tracking by APs}}
The AP performs digital beamforming to simultaneously form multiple beams. To avoid the inter-beam interference, different beams are allocated with orthogonal sensing spectrum bands~\cite{unified_ISAC_resource_allocation, Jie_chen_dual}, each having the bandwidth $B_{\rm s,0}$. For the AP, a constant number of $\widehat{X}_{\rm s}^{\rm A}$ spectrum bands are reserved to form $\widehat{X}_{\rm s}^{\rm A}$ search beams for identifying the targets that have newly arrived or become active within the RoIs of the mobile devices in its coverage~\cite{beam_allocation_search_beam_TSP}. In addition, $X_{\rm s}^{\rm A}$ spectrum bands are reserved for each AP to form $X_{\rm s}^{\rm A}\in \mathbb{Z}_{\geq 0}$ track beams, where $X_{\rm s}^{\rm A}$ is an optimization variable. Each reserved spectrum band can be used to track one associated target~\cite{Jie_chen_dual, beam_allocation_search_beam_TSP}. The total amount of spectrum resources reserved for target detection and tracking by the $L$ APs, denoted by $Z_{\rm s}^{\rm A}$, is $Z_{\rm s}^{\rm A} = L\cdot(\widehat{X}_{\rm s}^{\rm A}+X_{\rm s}^{\rm A})\cdot B_{\rm s,0}$. 

\subsubsection{Target Tracking by Mobile Devices}
\begin{figure}
    \centering
    \includegraphics[width=6.5cm]{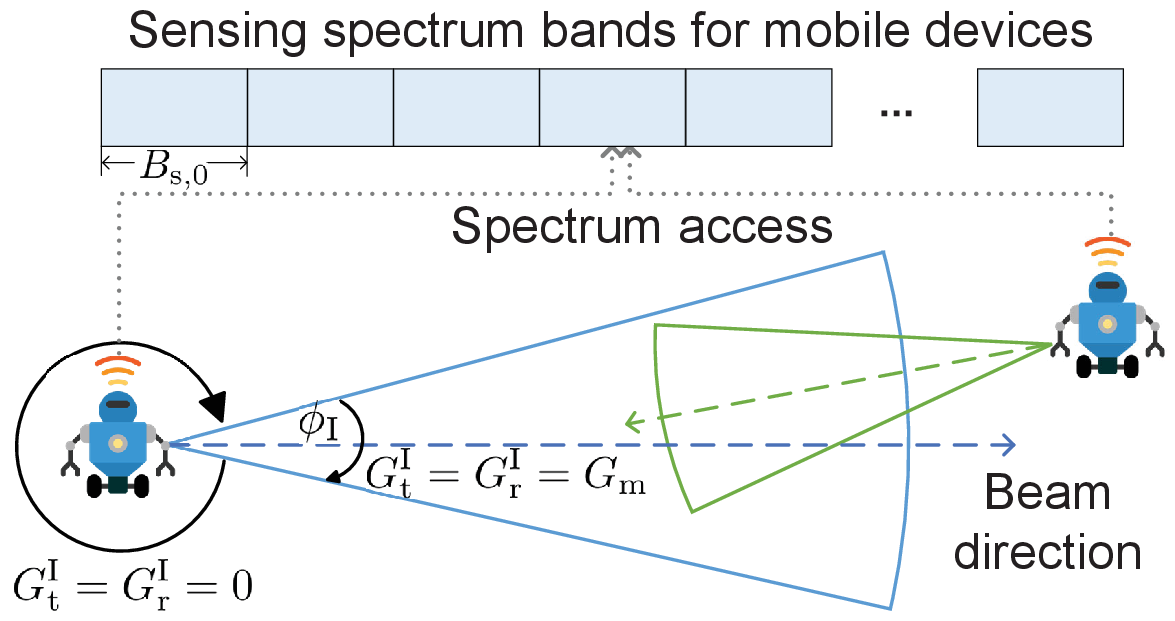}
    \caption{{The considered beam pattern of mobile devices and an illustration of the scenario where two mobile devices in the sensing mode interfere each other.}}
    \label{antenna_pattern}
\end{figure}

For tracking the targets by the mobile devices in the coverage of the $L$ APs, the global controller reserves $X_{\rm s}^{\rm I}\in \mathbb{Z}_{\geq 0}$ spectrum bands, each having bandwidth $B_{\rm s,0}$, where $X_{\rm s}^{\rm I}$ is an optimization variable. As a result, the total amount of spectrum resources reserved for target tracking by all the mobile devices is $Z_{\rm s}^{\rm I} = X_{\rm s}^{\rm I}\cdot B_{\rm s,0}$. At the beginning of a time slot, a mobile device will enter the sensing mode if not scheduled by the local controller for communication, the probability of which is $1-\rho_{\rm c}$. 
As shown in Fig.~\ref{antenna_pattern}, similar to~\cite{sensing_uncoordinated_SG}, in each time slot, the mobile device in the sensing mode randomly chooses one of the $X_{\rm s}^{\rm I}$ reserved spectrum bands with equal probability $1/X_{\rm s}^{\rm I}$, and forms a track beam towards one assigned target~\cite{ISAC_scheduling_yuguang}.\looseness=-1

As shown in Fig.~\ref{antenna_pattern}, we consider the following beam pattern of mobile devices in the sensing mode. The width of the beam is $\phi_{\rm I}$, and the antenna gain is $G_{\rm m}=2\pi G_0 /\phi_{\rm I} $ in the beam, and zero elsewhere~\cite{sector_gain_1}. The power of the signal reflected from a target received at the mobile device, denoted by $P_{\rm e}$, can be calculated by
$P_{\rm e} = {P_{\rm s} G_{\rm t}^{\rm I} G_{\rm r}^{\rm I} c^2\bar{{\sigma}}}/{(4\pi)^3f_{\rm s}^2 D^{4}}$~\cite{Letter_stochastic_radar_network}, where $G_{\rm t}^{\rm I}$ and $G_{\rm r}^{\rm I}$ both equal $G_{\rm m}$; $P_{\rm s}$ represents the transmit power of sensing signals; $c$ represents the speed of light; $\bar{\sigma}$ represents the average value of radar cross section (RCS) of the target; $f_{\rm s}$ represents the centre frequency of the accessed spectrum band; $D$ represents the distance between the mobile device and the target. 
{The mobile device in the sensing mode will receive interference from any other mobile device that is also in the sensing mode in the current time slot, accessing the same spectrum band for sensing, and residing within each other's beams.} For tractability, we consider only the strongest interfering signal to the mobile device, which is the signal from the nearest mobile device. The power of the interference signal is denoted by $I_{\rm s}$, and the signal-to-interference ratio (SIR) is denoted by $\gamma_{\rm s}$ and calculated as $\gamma_{\rm s} = {P_{\rm e}}/{I_{\rm s}}$.
\subsubsection{Sensing Data Processing}
{At the end of each sensing interval, the accumulated sensing data samples for each target in this interval are processed to update the target's state.} The total computational overhead for processing the sensing data of one target is denoted by $C_0$ in CPU cycles. The constant computational capability of each mobile device in CPU cycles per second is denoted by $F^{\rm I}_{\rm e}$. {For processing the sensing data collected by each AP, the amount of reserved edge computing resources at the AP is denoted by $F^{\rm A}_{\rm e}$, which is an optimization variable.} The total amount of edge computing resources reserved for the $L$ APs is $Z_{\rm e}^{\rm A} = L \cdot F^{\rm A}_{\rm e}$. The delay in processing the sensing data collected by the mobile device and the AP within a sensing interval, denoted by $T_{\rm e}^{\rm I} $ and $T_{\rm e}^{\rm A}$, respectively, is calculated as
$T_{\rm e}^{\rm I}  = {N^{\rm U}_{\rm I}C_0}/{F^{\rm I}_{\rm e}}$, and $T_{\rm e}^{\rm A} = {N^{\rm U}_{\rm A}C_0}/{F^{\rm A}_{\rm e}}$. Here, $N^{\rm U}_{\rm A}$ and $N^{\rm U}_{\rm I}$ represent the numbers of targets that are assigned to the AP and the mobile device at the beginning of a sensing interval.

\subsubsection{Sensing Requirements}
{Multi-dimensional requirements, including the quality, volume, and processing time of sensing data, are considered for the sensing service.

Firstly, the sensing data should have a high quality to provide accurate information of the targets. We use the SIR to assess the data quality, since it has been shown to significantly influence the accuracy of target state estimation~\cite{ten_challenges_JIoT} and target activity recognition~\cite{ambient_sensing}. 
Consider the randomness in the SIR across different sensing data samples, we adopt the constraint in a probabilistic form~\cite{TVT_stochastic_interference}:
$\mathbb{P}\{\gamma_{\rm s} \ge \widehat{\gamma}_{\rm s}\} \ge \widehat{P}$, where $\widehat{\gamma}_{\rm s}$ is the SIR threshold, and $\widehat{P}$ is the required minimum probability that the actual SIR meets or exceeds the SIR threshold.}

With an increasing distance between a mobile device and a target, the power of the received sensing signal from the target decreases, which leads to a higher chance of violating the SIR requirement. 
To satisfy the SIR requirement, the radius of the on-device sensing subregion needs to be properly determined, and the expected number of targets that can be assigned with the mobile device should satisfy the constraint:
\begin{equation}\label{SIR_constraint_transformed}
\begin{aligned}
\mathbb{E}\left[N^{\rm U}_{\rm I}\right] \le \mathbb{P}\left\{D^{\rm I, U}\le D^{\rm I,U}_{\rm max}\right\}\cdot \mathbb{E}\left[N^{\rm U}_{\rm I,1}\right],\\
\end{aligned}
\end{equation}
where $D^{\rm I, U}$ represents the distance between a target to its closest mobile device; $\mathbb{P}\{D^{\rm I, U}\le D^{\rm I,U}_{\rm max}\}$ is the probability that the distance is smaller than the radius of the on-device sensing subregion; $\mathbb{E}\left[N^{\rm U}_{\rm I,1}\right]$ is the expected number of targets that are closer to the mobile device than to any other mobile devices.

Secondly, the sensing data should have a large volume to provide sufficient information of the targets. We require that the average number of times each target is tracked per time slot should be higher than a threshold $\widehat{\rho}_{\rm s}\in(0,1]$~\cite{ISAC_scheduling}. Note that in each time slot, the expected number of tracking attempts that can be conducted by the mobile device and by the AP is $1-\rho_c$ and $X_{\rm s}^{\rm A}$ times, respectively. As a result, the sensing data volume requirement constrains the number of targets {that can be assigned to the mobile device and the AP, as follows:} 
\begin{equation}\label{num_targets_requirement_frequency}
\mathbb{E}\left[N^{\rm U}_{\rm I}\right] \le \frac{1-\rho_{\rm c}}{\widehat{\rho}_{\rm s}};~~
\mathbb{E}\left[N_{\rm A}^{\rm U}\right] \le \frac{X_{\rm s}^{\rm A}}{\widehat{\rho}_s}. 
\end{equation}

{Thirdly, the processing time of the sensing data should be low to ensure the freshness of sensing results. We require that the expected time for the mobile device and the AP to process the sensing data collected in each interval, i.e., $\mathbb{E}\left[T_{\rm e}^{\rm I}\right]$ and $\mathbb{E}\left[T_{\rm e}^{\rm A}\right]$, should be less than the duration of a sensing interval, or equivalently, $T$ time slots:}
\begin{equation}\label{comput_delay_requirement}
\frac{\mathbb{E}\left[N^{\rm U}_{\rm I}\right]C_0}{F^{\rm I}_{\rm e}} \le \tau T; ~~\frac{\mathbb{E}\left[N^{\rm U}_{\rm A}\right]C_0}{F^{\rm A}_{\rm e}} \le \tau T.
\end{equation}

\subsubsection{Sensing Service Demand and Capacity}
The number of the targets in the RoIs in the coverage of the AP is denoted by $N^{\rm U}$. The expectation of $N^{\rm U}$ corresponds to the sensing service demand in the coverage of the AP. The demand is calculated by multiplying the expected number of mobile devices with that of the targets within the RoI for each mobile device:
\begin{equation}\label{expected_N_U}
\mathbb{E}\left[N^{\rm U}\right] = \mathbb{E}\left[N^{\rm I}\right] \cdot\left(\mathbb{P}\{D^{\rm I, U}\le \hat{D}\}\cdot \mathbb{E}\left[N^{\rm U}_{\rm I,1}\right]\right),
\end{equation}
where $\mathbb{P}\{D^{\rm I, U}\le \hat{D}\}$ is the probability that the distance between a target to the closest mobile device is smaller than the radius of the RoI of the mobile device.

Combining the constraint on sensing quantity in~\eqref{num_targets_requirement_frequency}, the constraint on sensing data processing time in~\eqref{comput_delay_requirement}, and the constraint due to the maximum number of targets in the on-device sensing subregion in~\eqref{SIR_constraint_transformed}, we derive the maximum values of $\mathbb{E}[N^{\rm U}_{\rm I}]$ and $\mathbb{E}[N^{\rm U}_{\rm A}]$ as
\begin{subequations}\label{upper_bound_expected_targets}
\begin{equation}\label{upper_bound_NIU}
\begin{aligned}
\overline{N}^{\rm U}_{\rm I}  \triangleq  \min\left\{\mathbb{P}\left\{D^{\rm I, U}\le D^{\rm I,U}_{\rm max}\right\}\cdot \mathbb{E}\left[N^{\rm U}_{\rm I,1}\right],\frac{1-\rho_{\rm c}}{\widehat{\rho}_{\rm s}}, \frac{\tau T\cdot F^{\rm I}_{\rm e}}{C_0} \right\},
\end{aligned}
\end{equation}
\begin{equation}\label{upper_bound_NAU}
\begin{aligned}
\overline{N}_{\rm A}^{\rm U}\triangleq \min\left\{\frac{X_{\rm s}^{\rm A}}{\widehat{\rho}_s}, \frac{\tau T\cdot F^{\rm A}_{\rm e}}{C_0}\right\},
\end{aligned}
\end{equation}
\end{subequations}
which correspond to the sensing service capacity of each mobile device and each AP, respectively.
\subsection{Problem Formulation}
Our objective is to satisfy the {demands} for the sensing and communication services in the upcoming planning window with the minimal resource consumption. To achieve this objective, a decision $\Phi = \{D^{\rm I,U}_{\rm max}, \rho_{\rm c},X_{\rm c}^{\rm A}, X_{\rm s}^{\rm I}, X_{\rm s}^{\rm A},F^{\rm A}_{\rm e}\}$ is made, which consists of the radius of the on-device sensing subregion, i.e., $D^{\rm I,U}_{\rm max}$, the active probability for communication of each mobile device, i.e., $\rho_{\rm c}$, the amount of reserved spectrum resources, i.e., $X^{\rm A}_{\rm c}$, $X^{\rm A}_{\rm s}$, and $X^{\rm I}_{\rm s}$, and the amount of reserved edge computing resources, i.e., $F^{\rm A}_{\rm e}$. An optimization problem is formulated as follows:%
\begin{subequations}\label{original_prob}
\begin{align}
{\rm(P1):}\bm{\min}_{\Phi} ~~~& Z =\omega \left(Z_{\rm c}^{\rm A} +Z_{\rm s}^{\rm I} +Z_{\rm s}^{\rm A} \right) +\xi Z_{\rm e}^{\rm A}\label{P1obeject}\\
{\bm{\mathrm{s.t.~~~}}}& 0< \rho_{\rm c} \le 1,\label{P1constraint1}\\
&X_{\rm c}^{\rm A}, X_{\rm s}^{\rm I}, X_{\rm s}^{\rm A}\in \mathbb{Z}_{\geq 0},\label{P1constraint1}\\
& F^{\rm A}_{\rm e},D^{\rm I,U}_{\rm max} \in \mathbb{R}_{\geq 0} ,\label{P1constraint1}\\
&\bar{R}\ge \hat{R}, \label{comm_tot_requirement}\\
&\overline{N}_{\rm A}^{\rm U}\! +\! \mathbb{E}\left[N^{\rm I}\right]\!\cdot\!\overline{N}^{\rm U}_{\rm I}\!\ge\! \mathbb{E}\left[N^{\rm U}\right],\label{target_num_requirement_relaxed}\\
&\mathbb{P}\{\gamma_{\rm s} \ge \widehat{\gamma}_{\rm s}\} \ge \widehat{P}\label{SINR_constraint_in_problem},
\end{align}
\end{subequations}
where $Z$ is the overall resource consumption; $\omega$ and $\xi$ are the cost of reserving each unit of spectrum resources and edge computing resources, respectively; constraint \eqref{comm_tot_requirement} ensures that the expected transmission rate of each mobile device is greater than a threshold $\hat{R}$; constraint \eqref{target_num_requirement_relaxed} ensures that the expected number of targets that can be tracked with the requirements on the sensing data quantity and processing time satisfied is greater than the expected number of the targets in all the RoIs in the coverage of the AP; \eqref{SINR_constraint_in_problem} ensures that the requirement on sensing data quality evaluated by SIR is satisfied.

Solving Problem (P1) for network planning requires the spatial modeling of mobile devices and targets. In particular, according to~\eqref{low_bound_R_secIII}, calculating the communication service capacity, i.e., $\bar{R}$, requires the expected number of mobile devices in the coverage of each AP, i.e., $\mathbb{E}\left[N^{\rm I}\right]$; According to~\eqref{upper_bound_NIU}, the sensing capacity of each mobile device, i.e., $\overline{N}^{\rm U}_{\rm I}$, depends on the spatial distributions of targets; According to~\eqref{expected_N_U}, the sensing service demand in the coverage of each AP, i.e., $\overline{N}^{\rm U}$, depends on the spatial distributions of the mobile devices and the targets, respectively; Moreover, the sensing interference experienced by each mobile device, impacting $\gamma_{\rm s}$ in~\eqref{SINR_constraint_in_problem}, is affected by the spatial distribution of the mobile devices. {However, obtaining an accurate spatial model of the mobile devices and targets is challenging. This is because their spatial distributions can be non-stationary. For example, the safe separation distances between robots and human workers may change dynamically based on factors such as human and robot velocities and specific task requirements~\cite{marvel2017implementing}. Such non-stationarity can result in the drift of a spatial model for the spatial distributions and degrade the effectiveness of a network planning decision based on the spatial model.} 




\section{Network Slice DTs for Spatial Modeling and Network Planning}\label{chap_sliceDT}
\begin{figure}
    \centering
    \includegraphics[width=8cm]{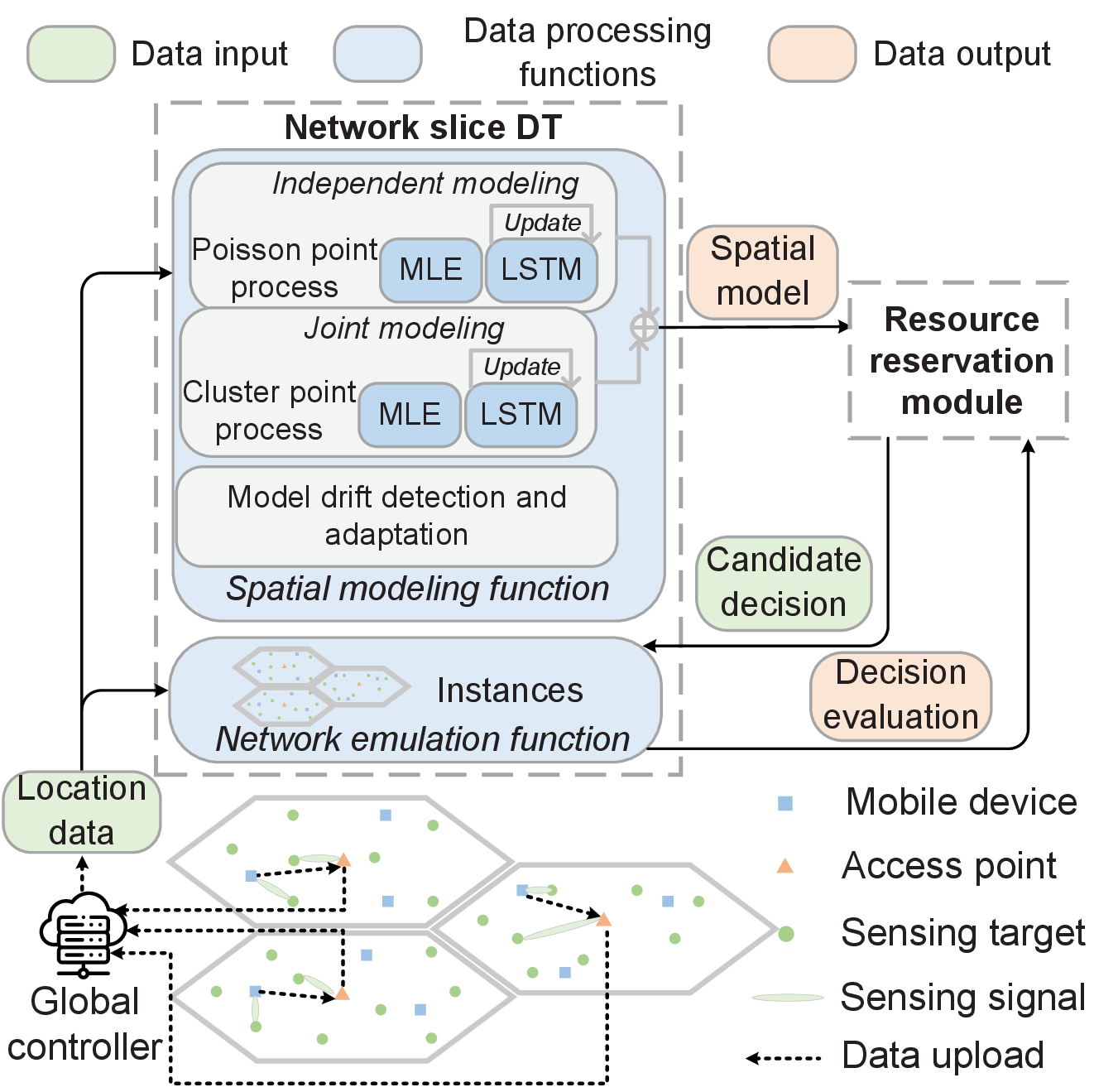}
    \caption{The DT of the sensing slice.}
    \label{slice_DT_model}
    \vspace*{-12pt}
\end{figure}
{The DT of a network slice, hereafter referred to as the \textit{slice DT}, is a digital representation of a network slice that is established and updated by collecting and processing data from the network slice for the service-specific prediction and control~\cite{shen2021holistic}. {We establish network slice DTs for adaptive spatial modeling and efficient and reliable network planning in cooperative ISAC networks. Particularly, in the slice DT corresponding to the sensing service, the spatial distributions of the mobile devices and targets are modeled, while in the slice DT corresponding to the communication service, only the spatial distribution of the mobile devices is modeled.} Note that we do not construct digital twins (DTs) for individual mobile devices and targets in spatial modeling, since it may incur significant overhead due to the potentially large number of entities and result in high network planning complexity.} 

{In Fig.~\ref{slice_DT_model}, we show the main components in the slice DT corresponding to the sensing service.} In this section, we first introduce the collection of the spatial distribution data. Then, we introduce a statistical spatial modeling function in the slice DTs based on two statistical spatial models with different modeling granularity. The spatial modeling function detects and adapts to the drift of individual statistical spatial models. Moreover, to ensure that a network planning decision can efficiently satisfy service demands, we propose an network emulation function in the slice DTs to efficiently evaluate a network planning decision.

\subsection{Spatial Distribution Data Collection}
To establish the slice DTs, each AP collects the spatial distribution data of the mobile devices and targets, and uploads them to the global controller every $M_0$ sensing intervals. Specifically, the locations of the targets assigned to the AP for tracking in this interval are obtained by processing the reflected sensing signals. In addition, the locations of the mobile devices and the targets assigned to the mobile devices are obtained and uploaded by the mobile devices to the AP. The spatial distribution data are used as the input for the slice DTs.

\subsection{{Statistical} Spatial Modeling Function}\label{subsec_spatial_modeling}
{At the end of the $k$-th planning window, the collected spatial distribution data of mobile devices and targets are processed for the statistical spatial modeling and network planning in the ($k+1$)-th planning window. To cope with the potential non-stationarity of the spatial distributions across planning windows, we adopt temporally non-stationary point processes~\cite{spatio_temporal_PP_review}.} Specifically, the model parameters are constant in each planning window but can change across planning windows. {To tackle the drift in predicting the model parameters, we synergize two statistical models with different numbers of model parameters.}

\subsubsection{Joint Spatial Modeling} In a joint spatial model, the spatial distributions of mobile devices and targets in the $k$-th planning window are jointly modeled by a Thomas cluster process. Specifically, the spatial distribution of mobile devices is characterized by a homogeneous Poisson point process (PPP) with intensity $\lambda_{k}^{\rm I}$. The relative two-dimensional locations of the targets clustering around each mobile device follows two identical and independent normal distributions with zero mean and standard deviation $\sigma_{k}^{\rm U}$~\cite{Thomas_contact_dist}. In addition, the number of targets in the cluster of each mobile device follows a Poisson distribution with mean $\mu_{k}^{\rm U}$. At the end of the $k$-th planning window, using {the spatial distribution data} collected in this planning window, { the reference values of the model parameters}, i.e., $\lambda_{k}^{\rm I}$, $\sigma_{k}^{\rm U}$, and $\mu_{k}^{\rm U}$, are determined based on the maximum likelihood estimation (MLE). {The prediction values of the three parameters} for network planning in the ($k+1$)-th planning window, denoted by $\check{\lambda}_{k+1}^{\rm I}$, $\check{\sigma}_{k+1}^{\rm U}$, and $\check{\mu}_{k+1}^{\rm U}$, are obtained by the long short-term memory (LSTM) neural networks based on the reference values of the parameters in the previous $K_0$ windows. 

\begin{figure}
    \centering
    \subfloat[$\mu_k^{\rm U}$ ]{\includegraphics[width=4.5cm]{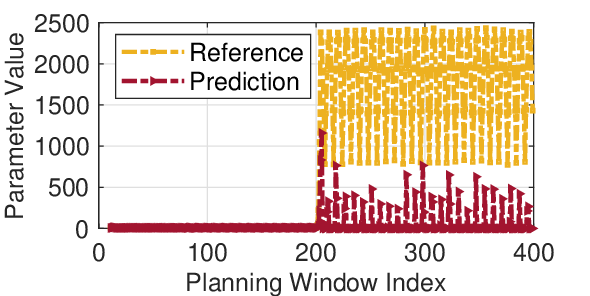}\label{parameter_subfigure_c}}
    \subfloat[$\sigma_k^{\rm U}$ ]{\includegraphics[width=4.5cm]{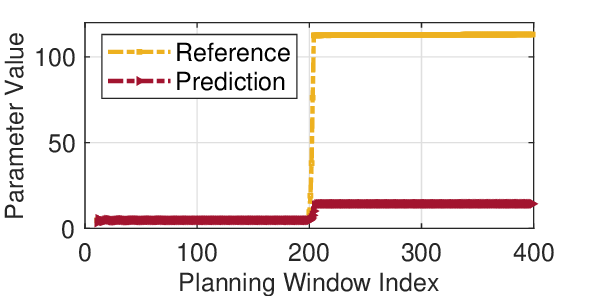}\label{parameter_subfigure_d}}
\caption{Values of the parameters $\mu_k^{\rm U}$ and $\sigma_k^{\rm U}$ in the joint spatial modeling across planning windows.}
\label{detailed_modeling_parameters}
\end{figure}

{The joint spatial modeling characterizes the spatial correlation between mobile devices and targets with the model parameters $\mu_k^{\rm U}$ and $\sigma_k^{\rm U}$. If the parameters can be accurately predicted for the upcoming planning window, the joint spatial modeling would be accurate. However, when the pattern of their spatial correlation changes, e.g., from attraction to repulsion, the values of the parameters that best fit the actual spatial distributions may exhibit a drastic change. In such a case, the accuracy of parameter prediction and that of the joint spatial modeling can degrade, referred to as {\textit{model drift}}.} To illustrate this, in Fig.~\ref{detailed_modeling_parameters}, we show the {reference} and prediction values of the model parameters $\mu_k^{\rm U}$ and $\sigma_k^{\rm U}$ in the joint spatial modeling, which are derived by MLE and the LSTM neural networks, respectively. The simulation settings are detailed in Section~\ref{simulation_sec}. It can be observed that after the $201$-st planning window, the {reference} values of \(\mu_k^{\rm U}\) and \(\sigma_k^{\rm U}\) exhibit a drastic increase, and the prediction values show an increasing gap with the {reference} values, indicating model drift.

\begin{figure}
    \centering
    \subfloat[$\lambda_k^{\rm I}$]{\includegraphics[width=4.5cm]{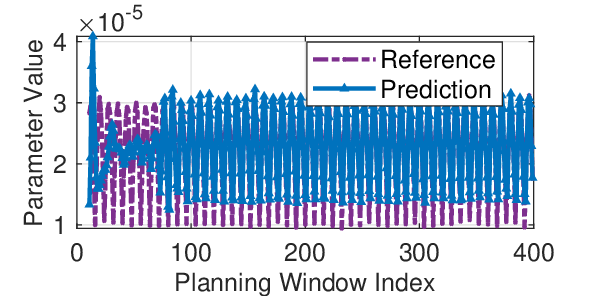}\label{parameter_subfigure_a}}
    \subfloat[$\lambda_k^{\rm U}$]{\includegraphics[width=4.5cm]{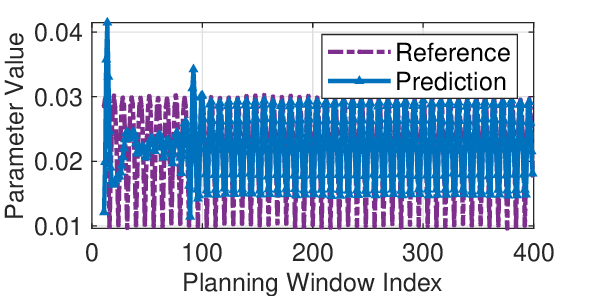}\label{parameter_subfigure_b}}
\caption{Values of the parameters $\lambda_k^{\rm I}$ and $\lambda_k^{\rm U}$ in the independent spatial modeling across planning windows.}
\label{simplified_modeling_parameters}
\end{figure}


\subsubsection{Independent Spatial Modeling}
To cope with the model drift, a spatial model robust to change in the spatial correlation of the mobile devices and targets is needed. We introduce the independent spatial modeling, which independently characterizes the spatial distributions of the mobile devices and targets. In particular, the locations of the mobile devices and targets in the $k$-th planning window are modeled by two \textit{independent} homogeneous PPPs with intensities $\lambda_{k}^{\rm I}$ and $\lambda_{k}^{\rm U}$, respectively. Similar to the joint spatial modeling, using the collected data in the $k$-th planning window, {the reference values of the intensity parameters in the two homogeneous PPPs}, i.e., $\lambda_{k}^{\rm I}$ and $\lambda_{k}^{\rm U}$, are determined based on the MLE. {The prediction values} of the two parameters for the $(k+1)$-th planning window, i.e., $\check{\lambda}_{k+1}^{\rm I}$, $\check{\lambda}_{k+1}^{\rm U}$, are obtained by LSTM networks based on reference values of the parameters in the previous $K_0$ windows.

{In Fig.~\ref{simplified_modeling_parameters}, for the same spatial distribution data as the joint spatial modeling in Fig.~\ref{detailed_modeling_parameters},} we show the {reference} and prediction values of all the parameters in the independent spatial modeling, i.e., $\lambda_{k}^{\rm I}$ and $\lambda_{k}^{\rm U}$. Note that the reference values evolve over planning windows in a manner similar to the absolute value of a cosine function, which makes the prediction at the lowest points of the reference values challenging and results in large discrepancies at those points. {Nevertheless, it can be observed that there is no drastic parameter value change and parameter prediction accuracy degradation since the $201$-st planning window. This observation suggests that the independent spatial modeling is more robust to model drift than the joint spatial modeling. However, it is important to note that in scenarios without model drift, the independent spatial modeling would yield lower accuracy than the joint spatial modeling due to its inability to characterize the spatial correlation between mobile devices and targets.}


\subsubsection{Model Drift Detection} This function monitors the prediction errors in the parameters of the spatial models to detect model drift. We develop a prediction error-based approach for model drift detection~\cite{model_drift_wireless}. Specifically, the errors in predicting ${\{\lambda}_{k}^{\rm I}, {\lambda}_{k}^{\rm U}, {\mu}_{k}^{\rm U}, {\sigma}_{k}^{\rm U}\}$ are evaluated by the mean absolute percentage error (MAPE). 
Let $H^{\rm S}_{k}$ and $H^{\rm D}_{k}\in\{0,1\}$ indicate whether the independent spatial modeling and the joint spatial modeling, respectively, experience model drift. At the end of the $k$-th planning window, the following three cases are considered for model drift detection. First, if the prediction of the parameter $\lambda^{\rm I}_{k}$, {which is shared by both spatial modelings}, exhibits drastic accuracy degradation, we set $H^{\rm S}_{k}=1$ and $H^{\rm D}_{k}=1$. Second, if the prediction of the parameter $\lambda^{\rm U}_{k}$, {which is unique to the independent spatial modeling}, exhibits drastic accuracy degradation, we set $H^{\rm S}_{k}=1$. Third, if parameters $\mu^{\rm U}_{k}$ and $\sigma^{\rm U}_{k}$, {which are unique to the joint spatial modeling}, exhibit exhibits drastic accuracy degradation, we set $H^{\rm D}_{k}=1$.
%

\subsubsection{Model Drift Adaptation} In this function, the following two approaches are designed to adapt to any model drift that is detected: (i) \textit{Model Update:} when model drift is detected, LSTM neural networks for predicting the parameters corresponding to the model drift are updated. In particular, in the planning windows after the detection, the {reference} values of the parameters will be collected and used to retrain the neural networks;
(ii) \textit{Model Ensemble:} when model drift is detected, the ensemble learning technique is exploited, and an ensemble spatial model that incorporates both the independent and the joint spatial models, is output by the model drift adaptation function. {In the absence of model drift, only the joint spatial model is output, since it offers potentially higher accuracy compared to the independent model and lower complexity for network planning compared to the ensemble model.}
\subsection{Network Emulation Function} \label{subsec_emulation_function}
{It is infeasible to directly fuse the joint and independent spatial models in the ensemble spatial model for network planning. Instead, we can evaluate and select from the two \textit{candidate} network planning decisions that are respectively obtained using the two spatial models. For the evaluation, we introduce a network emulation function that consists of the following two steps:} (i) \textit{Instance Construction:} multiple instances of both the sensing and communication slices are created. Each instance of a communication slice involves {deterministically distributed} APs and mobile devices, while each instance of a sensing slice involves {deterministically distributed} APs, mobile devices, and targets. 
For example, 
the locations of all the mobile devices and targets in the RoIs of the mobile devices at the end of the $M_0$-th sensing interval in the $k$-th planning window, along with the locations of the APs, are used to create {one} instance of the sensing slice; 
(ii) \textit{Decision Evaluation:} a candidate network planning decision is implemented on each of the instances. Then, the service demand and capacity of each instance are measured to evaluate the corresponding candidate decision. The measurement approach and the evaluation metric for a network planning decision are described in Section~\ref{subsec_emulation_decision_eval}.


\section{DT-Based Network Planning Algorithm}\label{overall_scheme_section}
{In this section, based on the statistical spatial models and the network emulation functions in the slice DTs, we propose a network planning algorithm.} 

\subsection{{ Statistical Spatial Modeling-Based Decision Making}}\label{subsec_decision_making}
In this subsection, we first derive the formulas for estimating the service demands and capacities in Problem (P1). Then, we analyze Problem (P1) for transforming the problem and developing an efficient algorithm to obtain a candidate network planning decision for the ($k+1$)-th planning window. {For brevity, we omit the subscript $k+1$ in the notation of the parameters that are predicted by LSTM neural networks.}
\subsubsection{Estimation of the Communication Service Capacity}
In both independent and joint spatial models, the mobile devices are modeled by a homogeneous PPP with the intensity $\check{\lambda}^{\rm I}$. Given that at least one mobile device exists, the expected number of mobile devices in the coverage of the AP with the area $A_0$ is given by
$\mathbb{E}\left[N^{\rm I}\right] = {\check{\lambda}^{\rm I}A_0}/({1-e^{-\check{\lambda}^{\rm I} A_0}}).$
By substituting $\mathbb{E}\left[N^{\rm I}\right]$ in~\eqref{low_bound_R_secIII} with this expression, the communication service capacity $\bar{R}$ in~\eqref{low_bound_R_secIII} can be calculated as
\begin{equation}\label{low_bound_R_secV}
\bar{R}= \frac{\rho_{\rm c}R_0X^{\rm A}_{\rm c}}{1+\left(\frac{\check{\lambda}^{\rm I}A_0}{\left(1-e^{-\check{\lambda}^{\rm I} A_0}\right)}-1\right)\rho_{\rm c}}.
\end{equation}

\subsubsection{Estimation of the Sensing Service Demand and Capacity} 
According to~\eqref{expected_N_U}, estimating the sensing demand in the coverage of each AP, i.e., $\mathbb{E}\left[N^{\rm U}\right]$, requires (i) the probability that a target is located in the RoI of the nearest mobile device, i.e., $\mathbb{P}\{D^{\rm I, U}\le \hat{D}\}$, and (ii) the expected number of targets closest to the mobile device, i.e., $\mathbb{E}\left[N^{\rm U}_{\rm I,1}\right]$.

First, the cumulative distribution function of the distance between a target to the closest mobile device, i.e., $D^{\rm I, U}$, is denoted by $\mathbb{P}\{D^{\rm I, U}\le d^{\rm I,U}\}$ and approximated as follows.
Given the independent spatial model, according to the contact distance distribution of a homogeneous PPP in~\cite{Jiang_hai_tutorial}, we have
$\mathbb{P}\{D^{\rm I, U}\le d^{\rm I,U}\}= 1\!-\!\exp\left[ -\!\left(d^{\rm I,U}\right)^2\pi\check{\lambda}^{\rm I}\right]$. Given the joint spatial model, according to the distribution of the distance in any cluster of a Thomas cluster process in~\cite{Thomas_contact_dist}, we have
$\mathbb{P}\{D^{\rm I, U}\le d^{\rm I,U}\} = 1\!-\!\exp\left[ -\!{\left(d^{\rm I,U}\right)^2}/({2(\check{\sigma}^{\rm U})^2})\right]$. Accordingly, $\mathbb{P}\{D^{\rm I, U}\le \hat{D}\}$ can be calculated.

Second, the expected number of targets closest to the mobile device, i.e., $\mathbb{E}\left[N^{\rm U}_{\rm I,1}\right]$, is approximated as follows.
Given the independent spatial model, $\mathbb{E}\left[N^{\rm U}_{\rm I,1}\right]$ is approximated as $\mathbb{E}\left[N^{\rm U}_{\rm I,1}\right] = {\check{\lambda}^{\rm U}}/{\check{\lambda}^{\rm I}}$, which is the ratio of the intensity of the targets to that of the mobile devices, due to the identical randomness across different mobile devices. Given the joint spatial model, $\mathbb{E}\left[N^{\rm U}_{\rm I,1}\right]$, is approximated as $\mathbb{E}\left[N^{\rm U}_{\rm I,1}\right] =  \check{\mu}^{\rm U}$, which is the expected number of targets in the cluster formed by the mobile device. According to~\eqref{upper_bound_NIU}, estimating the sensing capacity of each mobile device, i.e., $\overline{N}^{\rm U}_{\rm I}$, requires $\mathbb{P}\left\{D^{\rm I, U}\le D^{\rm I,U}_{\rm max}\right\}$ and $\mathbb{E}\left[N^{\rm U}_{\rm I,1}\right]$, which have been derived. The sensing capacity of each AP, i.e., $\overline{N}_{\rm A}^{\rm U}$, is not affected by the spatial model, and has been derived in \eqref{upper_bound_NAU}.


\subsubsection{Problem Analysis}\label{problem_and_closed_form}
We analyze the properties of Problem (P1) as follows.

{\noindent\textbf{Proposition 2:} \textit{In at least one optimal solution of Problem (P1), the radius of the on-device sensing subregion, i.e., $D^{\rm I,U}_{\rm max}$, is given by}
\begin{equation}\label{max_range}
\begin{aligned}
D^{\rm I,U}_{\rm max} &= \left(\frac{\bar{{\sigma}}}{\widehat{\gamma}_s}\cdot(-\ln \widehat{P})\cdot \frac{1}{\phi_{\rm I} ^2 \check{\lambda}^{\rm I}}\right)^{\frac{1}{4}}\cdot\left(\frac{X_{\rm s}^{\rm I}}{1-\rho_{\rm c}}\right)^{\frac{1}{4}},
\end{aligned}
\end{equation}
\textit{which is also the maximum distance between a target and an mobile device for satisfying the sensing SIR requirement in~\eqref{SINR_constraint_in_problem}.}}
\begin{proof}
Please refer to Appendix B.
\end{proof}
Based on Proposition 2, we can transform constraint~\eqref{SINR_constraint_in_problem} equivalently using~\eqref{max_range}. To make Problem (P1) more tractable, we relax the integer decision variables, i.e., ${X}^{\rm I}_{\rm s}, {X}^{\rm A}_{\rm c}, {X}^{\rm A}_{\rm s}$, to real-valued variables. Then, to efficiently solve the problem, we derive necessary conditions for optimality as follows.

\noindent\textbf{Proposition 3:} \textit{The optimal solution of Problem (P1) must satisfy the following three conditions:}
\begin{subequations}\label{necessary_opt_con}
\begin{equation}\label{rho_c_new_constraint}
\frac{1-\rho_{\rm c}}{\widehat{\rho}_{\rm s}} \le\min\left\{\mathbb{E}\left[N^{\rm U}_{\rm I,1}\right],\frac{\tau T\cdot F^{\rm I}_{\rm e}}{C_0}\right\};
\end{equation}
\vspace{-0.4cm}
\begin{equation}\label{rho_c_XIs_new_constraint}
\frac{1-\rho_{\rm c}}{\widehat{\rho}_{\rm s}} = \mathbb{P}\left\{D^{\rm I, U}\le D^{\rm I,U}_{\rm max}\right\}\cdot \mathbb{E}\left[N^{\rm U}_{\rm I,1}\right];
\end{equation}
\vspace{-0.4cm}
\begin{equation}\label{XsA_FeA_constraint}
\frac{X_{\rm s}^{\rm A}}{\widehat{\rho}_s}= \frac{\tau T\cdot F^{\rm A}_{\rm e}}{C_0}.
\end{equation}
\end{subequations}
\begin{proof}
Please refer to Appendix C.
\end{proof}
Condition \eqref{rho_c_new_constraint} specifies a minimum value of $\rho_c$ since the right-hand side is independent of any decision variable in Problem (P1); \eqref{rho_c_XIs_new_constraint} gives the relation between $\rho_c$ and ${X}^{\rm I}_{\rm s}$ since $\mathbb{P}\left\{D^{\rm I, U}\le D^{\rm I,U}_{\rm max}\right\}$ is a function of only these two decision variables; \eqref{XsA_FeA_constraint} gives the relation between ${X}^{\rm A}_{\rm s}$ and $F^{\rm A}_{\rm e}$. \looseness=-1
\subsubsection{Problem Transformation and Solution}
Based on \textbf{Proposition 2} and \textbf{Proposition 3}, given the value of the decision variable $\rho_{\rm c}$, the optimal values of the other four decision variables can be determined in closed forms. First, the minimum spectrum reservation for communication, ${X}^{\rm A}_{\rm c}$, can be obtained by converting the inequality constraint \eqref{comm_tot_requirement} to an equality constraint. Second, the optimal amount of spectrum resources reserved for sensing by mobile devices, ${X}^{\rm I}_{\rm s}$, can be found by solving \eqref{rho_c_XIs_new_constraint}. Third, the optimal spectrum reservation for sensing by APs, ${X}^{\rm A}_{\rm s}$, can be obtained by converting the inequality constraint \eqref{target_num_requirement_relaxed} to an equality constraint. Given the optimal ${X}^{\rm A}_{\rm s}$, the optimal edge computing network planning, $F^{\rm A}_{\rm e}$, can be obtained based on \eqref{XsA_FeA_constraint}. As a result, the original multi-variable optimization problem can be transformed into a single-variable problem concerning $\rho_{\rm c}$. The transformed problem can be solved by a brute-force search. Then, the real values in the decision are rounded up to integers. Given the joint spatial model or the independent spatial model, a candidate network planning decision can be respectively obtained.

\subsection{Network Emulation-Based Decision Evaluation}\label{subsec_emulation_decision_eval}
Based on the network emulation function of the slice DTs in Section~\ref{subsec_emulation_function}, a candidate network planning decision, denoted by $\Phi$, can be evaluated in the following way.
\subsubsection{Evaluation under Communication Service Provision}
Evaluating the \textit{average} transmission rate of mobile devices requires network emulation over a long period. To reduce the complexity, we use the \textit{expectation} of the transmission rate derived in~\eqref{low_bound_R_secV}. Given the spatial distribution data of mobile devices in all the constructed instances, the intensity of mobile devices is determined by MLE and used to replace the one predicted by an LSTM network, i.e., $\check{\lambda}^{\rm I}$ in~\eqref{low_bound_R_secV}. Then, given the value of $\rho_c$ in a network planning decision $\Phi$, the lower bound of the expected achievable transmission rate, i.e., $\bar{R}$, is calculated as the network emulation-based estimation of the communication service capacity. The evaluation metric chosen here is the relative difference between the service demands and capacities. Specifically, for the communication service, the relative difference is denoted by $\Delta_{\rm c} (\Phi)$ and calculated as:
$\Delta_{\rm c} (\Phi) = {|\bar{R}-\hat{R}|}/{\hat{R}}$, which can reflect both the resource over-provision, where the communication capacity exceeds the communication demand, i.e., $\bar{R} > \hat{R}$, and resource under-provision, i.e., $\bar{R} < \hat{R}$.

\subsubsection{Evaluation under Sensing Service Provision}
Similarly, given the candidate network planning decision $\Phi$, the number of the mobile devices, the number of targets that can be monitored by the mobile devices, and the number of targets that reside in the RoIs of all the mobile devices in the network instances are obtained and averaged as the network emulation-based estimation of $\mathbb{E}[N^{\rm I}]$, $\overline{N}_{\rm I}^{\rm U}$, and $\overline{N}^{\rm U}$, respectively. The relative difference between the sensing demand and sensing capacity is denoted by $\Delta_{\rm s}(\Phi)$ and calculated as
$\Delta_{\rm s} (\Phi) = {\left|\overline{N}^{\rm U} - \left(\overline{N}_{\rm A}^{\rm U} + \mathbb{E}\left[N^{\rm I}\right]\cdot\overline{N}^{\rm U}_{\rm I}\right)\right|}/{\overline{N}^{\rm U}}$.
\subsubsection{Overall Evaluation of a Network Planning Decision}
Given a candidate network planning decision $\Phi$, the average difference between the demands and capacities for the sensing and communication services is denoted by $\Delta(\Phi)$ and calculated as $\Delta(\Phi) = {(\Delta_{\rm c} (\Phi)+\Delta_{\rm s}(\Phi))}/{2}$, which is used for the overall evaluation of the network planning decision.

\subsection{Overall DT-Based Network Planning Algorithm}
Based on the proposed closed-form candidate network planning decision making and the network emulation-based decision evaluation approaches, we propose a network planning algorithm that can adapt to the drift of individual spatial models as follows: (i) \textit{Case 1 (Model Drift Not Detected):} When model drift is not detected, i.e., $H^{\rm S}_{k}=0$ and $H^{\rm D}_{k}=0$, only the joint spatial model is output from the slice DT. The network planning decision based on the joint spatial model, denoted by $\Phi^{\rm D}$, is obtained according to Section~\ref{subsec_decision_making}, and adopted for the ($k+1$)-th planning window. (ii) \textit{Case 2 (Model Drift Detected):} When model drift is detected for the joint modeling, i.e., $H^{\rm S}_{k}=1$ or $H^{\rm D}_{k}=1$, the slice DTs would output an ensemble spatial model consisting of both the joint and the independent spatial models. In this case, the two candidate network planning decisions are respectively derived as $\Phi^{\rm S}$ and $\Phi^{\rm D}$. Then, the two candidate decisions are respectively evaluated by the network emulation function according to Section~\ref{subsec_emulation_decision_eval}, and the evaluation results $\Delta(\Phi^{\rm S})$ and $\Delta(\Phi^{\rm D})$, reflecting the respective discrepancies between service demands and service capacities, are obtained. The candidate decision with a smaller demand-capacity discrepancy is selected as the final network planning decision for the ($k+1$)-th planning window.

\section{Simulation Results}\label{simulation_sec}
In this section, we demonstrate by simulations the performance of our proposed DT-based network planning algorithm for cooperative ISAC networks.

\subsection{Simulation Settings and Benchmarks}


\begin{table}[h!]
\centering
\caption{System parameters in simulation}
\label{tab:simulation_parameters}
\begin{tabular}{l| c}
\hline
\textbf{Parameter} & \textbf{Value} \\
\hline
Beamwidth of mobile devices for sensing, $\phi_{\rm I}$& $\pi/6$\\ 
Bandwidth of a communication subcarrier, $B_{\rm c,0} $ & $0.015$ MHz \\
Transmit power for sensing, $P_{\rm s}$ & $ 20$ dBm\\
Bandwidth of a sensing band, $B_{\rm s,0}$ & $1$ MHz \\
Average radar cross section, $\bar{\sigma}$  & $0.5$ $\rm{m}^2$ \\
Side length of each AP's coverage, $r_0$ & $500$ m\\
Duration of a time slot, $\tau$ & $1$ ms\\
Number of time slots in a sensing interval, $T$ & $1000$ \\
Number of sensing intervals in a planning window, $M$ & $500$ \\
Location data collection period in sensing intervals, $M_0$ & $10$ \\
Number of APs in the cluster, $L$ & $3$\\
CPU cycles for sensing one target, $C_0$ & $1\times 10^{8}$ cycles \\
CPU frequency of each mobile device, $F_{\rm e}^{\rm I}$ & $2$ GHz\\
SIR requirement for sensing, $\widehat{\gamma}_s$ & $20$ dB \\
Threshold for SIR requirement satisfaction, $\hat{P}$ & $0.95$ \\
Communication throughput requirement, $\hat{R}$ & $1$ kb/slot\\
Sensing frequency requirement for each target, $\hat{\rho}_{\rm s}$ & $0.05$ \\
Radius of the RoI for each mobile device, $\hat{D}$ & $10$ m\\
Cost of reserving a unit of spectrum resources, $\omega$ & $1$\\
Cost of reserving a unit of edge computing resources, $\xi$ & $1\times10^{-3}$\\
\hline
\end{tabular}
\vspace{-10pt}
\end{table}

{The synthetic spatial distributions of the mobile devices and targets in the simulations are generated in the following way.} The locations of mobile devices are generated following a homogeneous PPP, while the locations of targets are generated in two steps. In the first step, the number of targets in the RoI of each mobile device is determined by sampling from a Poisson distribution. In the second step, targets are randomly and uniformly generated within the RoI of each mobile device. A thinning approach is then used to determine whether to retain each target based on a probability, until the predetermined number of targets for each mobile device is reached. Specifically, two types of mobile devices are considered, with devices set to type I and type II with probabilities $\nu$ and $1-\nu$, respectively. The probabilities to retain a generated target for type I and type II mobile devices are given by $D^{\rm I, U}/\hat{D}$ and $1 - D^{\rm I, U}/\hat{D}$, respectively, where $D^{\rm I, U}$ is the distance between the target and the mobile device, and $\hat{D}$ is the radius of the RoI. {The spatial correlation pattern between the mobile devices and the sensing targets around them differs for the two types of mobile devices. When the value of $\nu$ changes, the proportions of the two types of the mobile devices and the overall spatial correlation pattern will be changed, which may result in the drift of joint spatial model which characterizes the spatial correlation.} The other main simulation settings are given in Table~\ref{tab:simulation_parameters}.

To generate the spatial distribution data across planning windows, the parameters of the homogeneous PPP and the Poisson distribution are changed based on the absolute values of cosine functions with the angular frequency $\pi/8$. 
Each simulation run includes $400$ consecutive planning windows, and for each simulation setting, we conduct $5$ simulation runs and evaluate the average performances. The following three performance metrics are adopted: (i) \textit{Actual Service Satisfaction Ratio:} this metric represents the average satisfaction ratios for the sensing and communication services as observed at the end of each planning window. Each service's satisfaction ratio is calculated as the minimum between $1$ and the ratio of the service capacity to the service demand; (ii) \textit{Overall Resource Consumption:} this metric represents the overall cost of reserving the spectrum resources and edge computing resources for the sensing and communication services, as defined in~\eqref{P1obeject}; (iii) \textit{Modeling Error}: this metric represents the average prediction error of the parameters in a spatial model. A sliding average approach is used with the width of $20$ planning windows.

We compare our proposed algorithm with the following benchmarks: (i) \textit{Independent Modeling (Ideal Case):} use the independent spatial modeling with perfect parameter prediction; (ii) \textit{Independent Modeling:} use the independent spatial modeling with LSTM neural networks to predict model parameters, while the proposed model drift detection and the model update approaches are not implemented; (iii) \textit{Joint Modeling (Ideal Case):} use the joint spatial modeling with perfect parameter prediction; (iv) \textit{Joint Modeling:} use the joint spatial modeling with LSTM neural networks to predict model parameters, while the proposed model drift detection and model update approaches are not implemented.

\subsection{Effectiveness of DT-Based Network Planning Algorithm}
\begin{figure}
    \centering
    \subfloat[Actual service satisfaction ratio.]{\includegraphics[width=6.5cm]{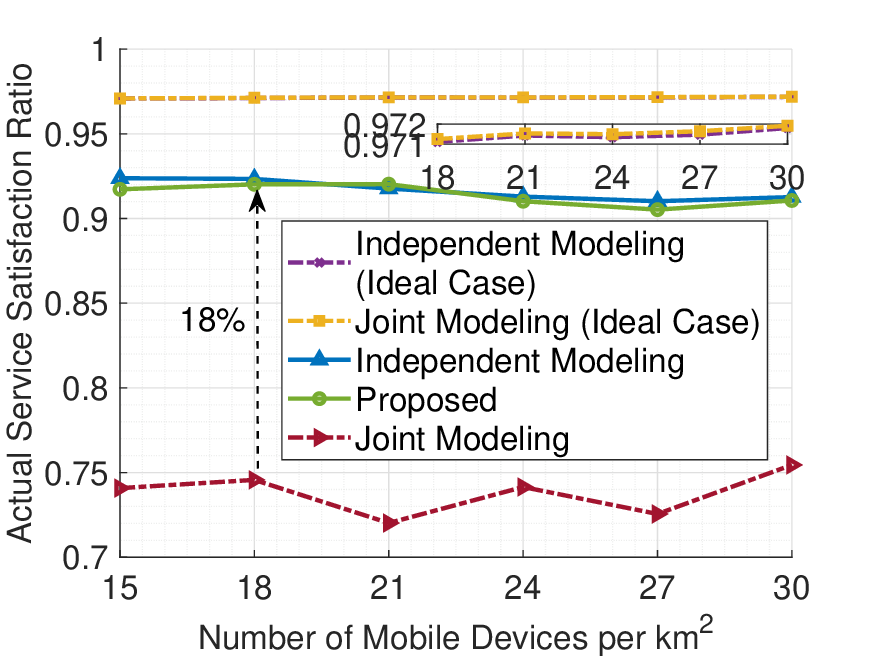}
    \label{Fig1_satisfaction_ratio}}
    \\
    \subfloat[Overall resource consumption.]{\includegraphics[width=6.5cm]{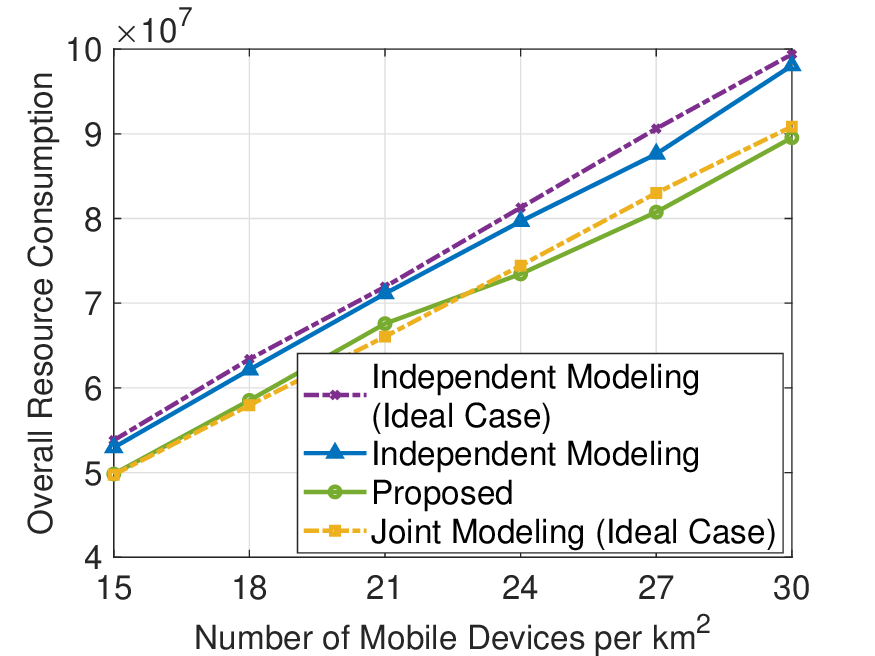}
    \label{resource_consumption}}
    \caption{Performance evaluation with varying number of mobile devices per square kilometre.}
    \label{Fig_performance_varying_density}
    \vspace*{-12pt}
\end{figure}

In this subsection, we demonstrate the performance of the DT-based network planning algorithm. In each simulation, the value of $\nu$ is set to $0.1$ and $0.7$ before and after the $201$-st planning window. The average number of targets in the RoI of each mobile device is set to be $5.5$ and that of the mobile devices varies from $15$ to $30$ per square kilometre. In Fig.~\ref{Fig1_satisfaction_ratio}, we show the actual service satisfaction ratio, and in Fig.~\ref{resource_consumption}, we show the overall resource consumption.


First, in the two ideal cases where parameter predictions are perfect, we compare the performances of the joint modeling and independent modeling. It can be observed from Fig.~\ref{Fig1_satisfaction_ratio} that the use of a single model in the ideal cases result in similarly high service satisfaction ratios. From Fig.~\ref{resource_consumption}, it can be observed that the use of the joint modeling in the ideal case results in the network planning decisions that consume less resources than the use of independent spatial modeling in the ideal case. {This is because, even when the independent modeling may accurately estimate the expected number of targets in the whole RoI of each mobile device, it underestimates the number in the on-device sensing subregion, since the spatial attraction between mobile devices and targets is not characterized.} Consequently, the demand for sensing services can be accurately estimated, whereas the sensing capacity of mobile devices is under-estimated. As a result, although the service demand can be satisfied, spectrum and edge computing resources would be over-provisioned.



Second, in the non-ideal cases where model parameters are predicted by LSTM neural networks, we compare the performances of the independent modeling, joint modeling, and our proposed DT-based network planning algorithms. Unlike in the ideal cases where the joint modeling outperforms the independent modeling, in the non-ideal cases, it can be observed from Fig.~\ref{Fig1_satisfaction_ratio} that, the joint modeling for network planning results in a much lower actual service satisfaction ratio than the independent modeling. This is because, as shown in Fig.~\ref{detailed_modeling_parameters}, after the value of $\nu$ changes in the $201$-st planning window, the joint spatial modeling exhibits model drift, while the independent spatial modeling does not. {Specifically, as shown in Fig.~\ref{parameter_subfigure_c}, the average number of targets clustering around each mobile device is under-predicted by the joint modeling, which results in the under-prediction of the sensing service demand. In addition, as shown in Fig.~\ref{parameter_subfigure_d}, the standard derivation of the distance between a target and a mobile device is under-predicted by the joint modeling, which results in the over-prediction of mobile devices' sensing capability.}
In addition, it can be observed from Fig.~\ref{Fig1_satisfaction_ratio} that using the DT-based network planning algorithm results in nearly the same service satisfaction ratio as the independent spatial modeling-based network planning. Nevertheless, it can be observed from Fig.~\ref{resource_consumption} that, the use of the DT-based network planning algorithm results in lower resource consumption compared with the independent spatial modeling-based network planning algorithm.

From the observations in the non-ideal cases, our proposed DT-based network planning algorithm achieves the best performance. 
In the following subsections, we illustrate the benefits of our proposed model update and model ensemble approaches, respectively, where the average number of targets in the RoI of each mobile device is $5.5$ and that of the mobile devices is $15$.


\subsection{Effectiveness of Model Update}
\begin{figure}
    \centering
    \includegraphics[width=6.5cm]{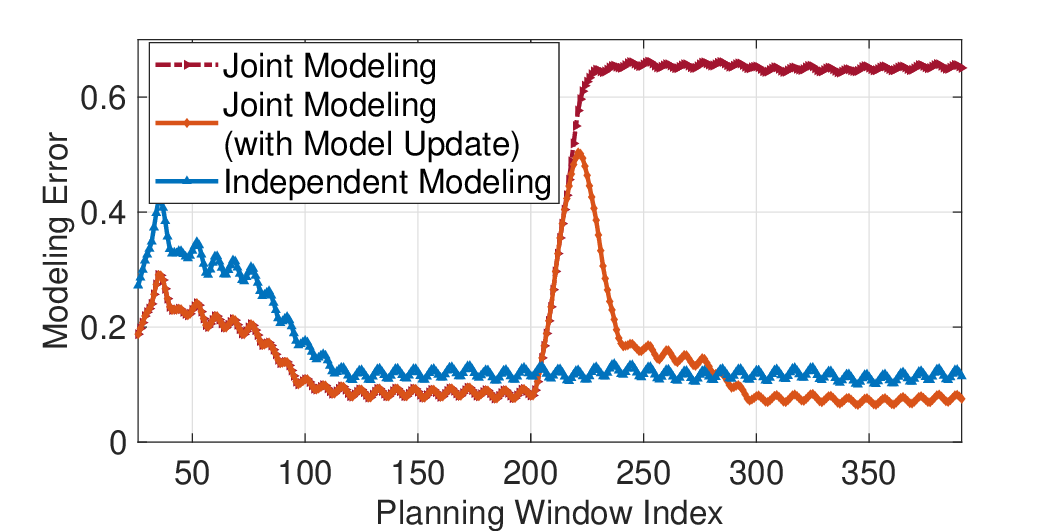}
    \caption{Modeling error across planning windows.}
    \label{Fig3_mape_prediction_error}
    \vspace*{-12pt}
\end{figure}

In Fig.~\ref{Fig3_mape_prediction_error}, we show the modeling error across planning windows. From Fig.~\ref{Fig3_mape_prediction_error}, it can be observed that,  after the value of $\nu$ changes in the $201$-st planning window, the modeling error of the joint modeling drastically increases, while that of the independent modeling does not. In contrast, with model drift detection and model update by retraining the LSTM neural networks, the modeling error of the joint modeling quickly falls down after the increase. {This observation demonstrates that our model update approach can enable the joint spatial modeling to tackle the model drift issue, making it suitable for network planning in the presence of model drift.}

\subsection{Effectiveness of Model Ensemble}
{In this subsection, we demonstrate the performance with the model drift of the joint spatial modeling occurring at different frequencies, which are defined as the total numbers of drift occurrences that are uniformly distributed across consecutive $400$ planning windows.} To simulate an occurrence of the model drift, the value of $\nu$ alters between $0.1$ and $0.7$. 

\begin{figure}
    \centering
    \subfloat[Model drift indicator $H^{\rm D}_k$.\label{drift_detection_performance}]{\includegraphics[width=6.5cm]{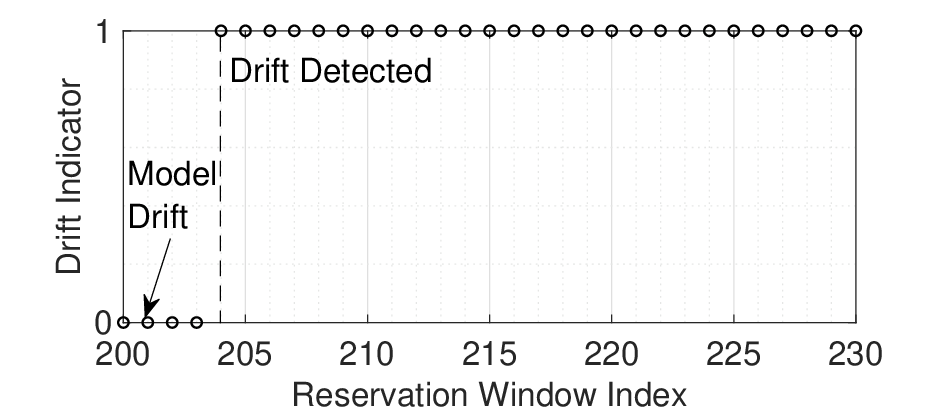}}
    \hfill
    \subfloat[Actual service satisfaction ratio.\label{average_satisfaction_snapshot}]{\includegraphics[width=6.5cm]{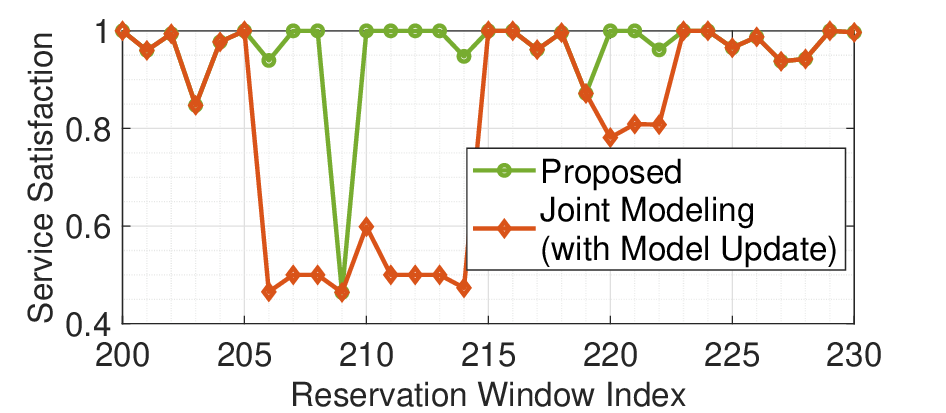}}
    \hfill
    \subfloat[Overall resource consumption.\label{average_resource_snapshot}]{\includegraphics[width=6.5cm]{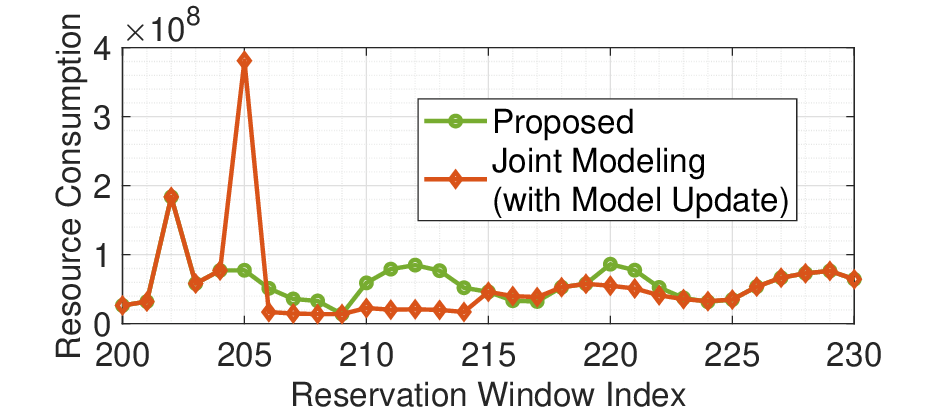}}
    \caption{Performance evaluation in the early planning windows after model drift occurs.}
    \label{snapshot_performance}
    \vspace*{-12pt}
\end{figure}

{First, in Fig.~\ref{snapshot_performance}, we show the per-window performances from the $200$-th to the $230$-th planning window, where drift occurs $3$ times in 400 windows. In Fig.~\ref{drift_detection_performance}, we show the model drift detection result. From Fig.~\ref{drift_detection_performance}, it can be observed that the model drift, occurring in the $201$-st planning window, is successfully detected at the $204$-th planning window.} In Fig.~\ref{average_satisfaction_snapshot}, we show the actual service satisfaction ratio. From the figure, it can be observed that, after the model drift occurs, the actual service satisfaction ratio, by using the joint modeling with the model update, first decreases, and then increases. 
In addition, it can be observed that by using the model ensemble approach in the DT-based adaptive modeling, a higher service satisfaction ratio can be achieved in multiple planning windows. This is because, by using the model ensemble approach, when model drift is detected, both the joint and independent spatial models are output from the slice DTs for network planning, and the network emulation function in the slice DTs evaluates the network planning decisions derived from the two spatial models, and selects the better decision. {Similarly, from Fig.~\ref{average_resource_snapshot}, it can be observed that, the model ensemble approach demonstrates its effectiveness by preventing resource over-provision in the $205$-th planning windows, while avoiding under-provision in the remaining windows between the $206$-th and $214$-th.}

\begin{figure}
    \centering
    \subfloat[Actual service satisfaction ratio.]{\includegraphics[width=6.5cm]{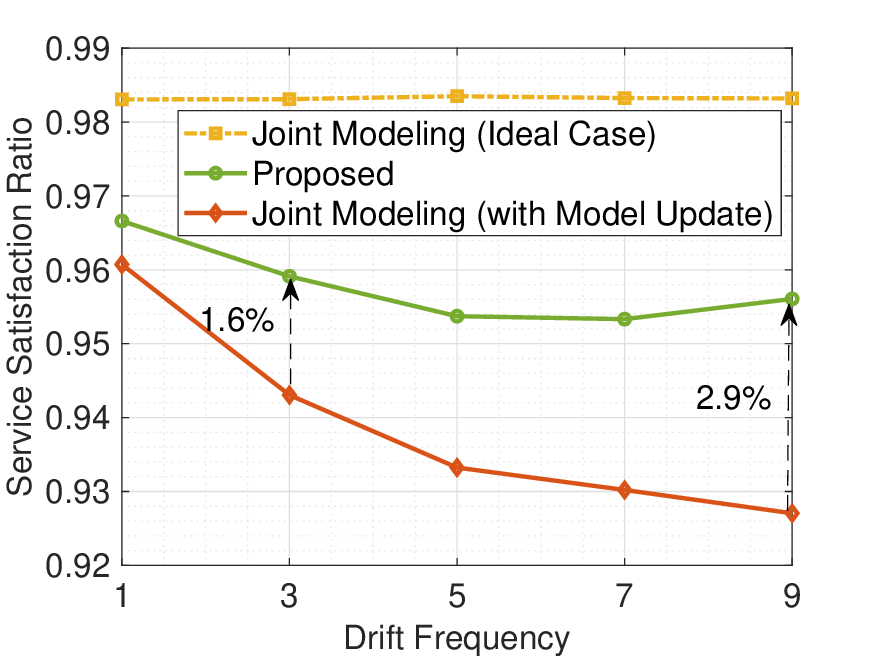}
    \label{Fig4_satisfactionratio_varying_drift_windows}}
    \hspace{1cm}
    \subfloat[Overall resource consumption.]{\includegraphics[width=6.5cm]{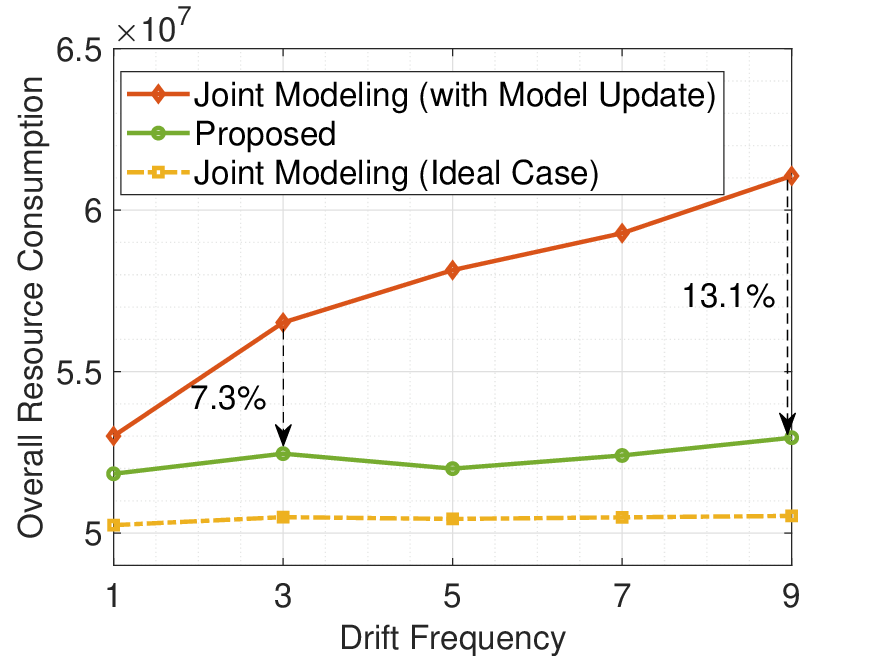}
    \label{Fig5_resourceconsumption_varying_drift_windows}}
    \hspace{1cm}
    \subfloat[Joint modeling utilization ratio.]{\includegraphics[width=6.5cm]{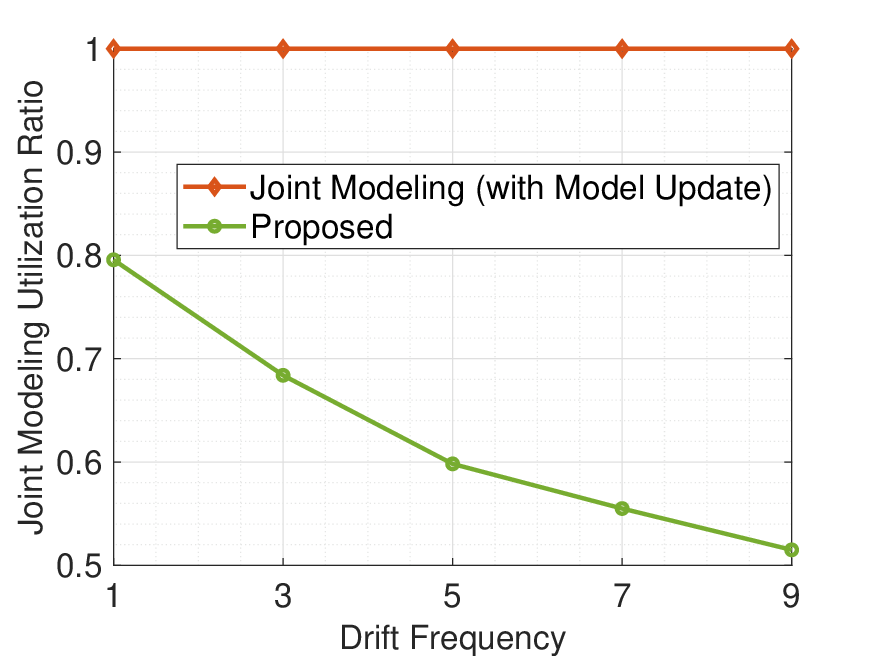}\label{utilization_ratio_drift_frequency}}
    \caption{Performance evaluation under varying model drift frequency.}
    \label{Fig_performance_varying_drift}
    \vspace*{-12pt}
\end{figure}

Second, we show the average performance across planning windows when the model drift frequency takes different values. {Note that we only consider odd integer values for the drift frequency because they result in equal proportions of the two types of mobile devices. Consequently, although the service provisions differ between the two types, the overall service provision performance in the ideal case remains the same across the different drift frequencies. This consistency helps clearly reveal the impact of the drift frequency on our proposed scheme and the benchmark schemes.} In Fig.~\ref{Fig4_satisfactionratio_varying_drift_windows} and Fig.~\ref{Fig5_resourceconsumption_varying_drift_windows}, we show the actual service satisfaction ratio and the overall resource consumption, respectively. From the two figures, it can be observed that compared with always using the joint modeling, using the model ensemble approach in our proposed DT-based spatial modeling function improves the actual service satisfaction ratio and decreases the overall resource consumption. 
In addition, with a higher model drift frequency, the advantage of DT-based spatial modeling function, in terms of improving the actual service satisfaction and reducing the resource consumption, increases. {This is because, with the increase of drift frequency, the joint modeling would result in degraded performance, i.e., reduced service satisfaction ratio and increased resource consumption. Nevertheless, in our proposed DT-based spatial modeling function, as shown in Fig.~\ref{utilization_ratio_drift_frequency}, the joint modeling utilization ratio, defined by the percentage of the planning windows when the joint spatial modeling is selected, decreases with the increase of the drift frequency. In this way, the service provision performance becomes less invariant to the increase of drift frequency, similar to ideal case as shown in Fig.~\ref{Fig4_satisfactionratio_varying_drift_windows} and Fig.~\ref{Fig5_resourceconsumption_varying_drift_windows}.} 

The observations in this subsection demonstrate that it is not always beneficial to characterize the spatial correlation between mobile devices and targets for resource management, {especially when the spatial correlation can experience drastic variations. Instead, by using our proposed DT-based spatial modeling and network emulation functions, the spatial correlation can be properly characterized for effective resource management in cooperative ISAC networks.}

\section{Conclusion}\label{sec_conclusion}
{In this paper, we have proposed a novel drift-adaptive slicing-based resource management scheme to support the efficient collaboration of mobile devices and APs in sensing and communication service provision.} We have developed DTs of the sensing and communication slices for drift-adaptive spatial modeling and efficient network emulation, {which empowers efficient and reliable slicing-based resource management decision-making in non-stationary environments.} In addition to network-level spatial modeling and emulation, our proposed DT design can also be applied to adaptively characterize the service demand and satisfaction of individual users. For future work, we will investigate the potential of user DTs in characterizing individual users' {diverse and dynamic} attention patterns towards various targets, enabling user-centric sensing service provision within ISAC paradigms.

\section*{Appendix A\\ Proof of Proposition 1}
If the mobile device is scheduled for communication in the slot, we have $R=X_{\rm c}^{\rm I}R_0$; and otherwise, $R=0$. As a result, the expected transmission rate for the device is
\begin{subequations}\label{commun_tot}
\begin{align}
\mathbb{E}_{N^{\rm I}_{\rm c}}\left[R\right] &=  \rho_{\rm c}\cdot\mathbb{E}_{N^{\rm I}}\left[\mathbb{E}_{N^{\rm I}_{\rm c}}\left[X_{\rm c}^{\rm I}R_0|N^{\rm I}\right]\right]\\
&\ge \rho_{\rm c}\cdot\mathbb{E}_{N^{\rm I}}\left[\frac{R_0X^{\rm A}_{\rm c}}{1+(N^{\rm I}-1)\rho_{\rm c}} \right]\label{comm_rate_inequality_1}\\
&\ge \rho_{\rm c}\cdot\frac{R_0X^{\rm A}_{\rm c}}{1+(\mathbb{E}\left[N^{\rm I}\right]-1)\rho_{\rm c}}\label{comm_rate_inequality_2},
\end{align}
\end{subequations}
where~\eqref{comm_rate_inequality_1} holds since, based on Jensen's inequality, we have
\begin{equation}
\mathbb{E}_{N^{\rm I}_{\rm c}}\left[X_{\rm c}^{\rm I}|N^{\rm I}\right] \ge \frac{X^{\rm A}_{\rm c}}{\mathbb{E}_{N^{\rm I}_{\rm c}}\left[N^{\rm I}_{\rm c}|N^{\rm I}\right]} = \frac{X^{\rm A}_{\rm c}}{1+(N^{\rm I}-1)\rho_{\rm c}},
\end{equation}
and~\eqref{comm_rate_inequality_2} holds also due to the Jensen's inequality.

\section*{Appendix B\\ Proof of Proposition 2}


Under both the independent modeling and the joint modeling, the targets are modeled to be uniformly distributed in various directions around each mobile device. As a result, the directions of the beams formed by mobile devices in the sensing mode are independently and uniformly distributed in $[0, 2\pi)$~\cite{Letter_stochastic_radar_network, TVT_stochastic_interference}. In addition, since mobile devices independently access each spectrum band for the sensing with the probability $(1-\rho_{\rm c})/X_{\rm s}^{\rm I}$, based on the property of PPPs, the mobile devices that access the same spectrum band for sensing form a thinned homogeneous PPP with the intensity
$\check{\lambda}^{\rm I}\cdot(1-\rho_{\rm c})/X_{\rm s}^{\rm I}$. According to~\cite[Proposition 1]{Letter_stochastic_radar_network}, the CDF of the strongest interference $I_{\rm s}$ for the mobile device is :
\begin{equation}
\begin{aligned}
\mathbb{P}\{I_{\rm s}\le i_{\rm s}\}
& = \exp\left(-\frac{\check{\lambda}^{\rm I}\cdot(1-\rho_{\rm c})}{X_{\rm s}^{\rm I}} \cdot \frac{\phi_{\rm I} ^2}{4\pi} \cdot \left(\frac{P_{\rm s}G_{\rm m}^2c^2}{(4\pi f_{\rm s})^2}\right)\cdot i_{\rm s}^{-1}\right).
\end{aligned}
\end{equation}
Since $\gamma_{\rm s} = P_{\rm e}/I_{\rm s}$, we have $\gamma_{\rm s}$ is $\mathbb{P}\{\gamma_{\rm s} \ge \widehat{\gamma_{\rm s}}\} = \mathbb{P}\{I_{\rm s}\le\frac{P_{\rm e}}{\widehat{\gamma_{\rm s}}}\}$, which should be greater than $\hat{P}$ to satisfy the sensing quality requirement. Therefore, we have
\begin{equation}\label{power_contraint}
\begin{aligned}
P_{\rm e} \ge \hat{\gamma}_{\rm s}\left(-\ln \hat{P}\right)^{-1}\cdot\left(\frac{\phi_{\rm I} ^2 (1-\rho_{\rm c})\lambda_1^{\rm I}} { 4\pi X_{\rm s}^{\rm I}}\right)\cdot\frac{P_{\rm s}G_{\rm m}^2c^2}{(4\pi f_{\rm s})^2}.
\end{aligned}
\end{equation}
The power received sensing signals, i.e., $P_{\rm e}$, is negatively correlated with the distance between the mobile device and the target, i.e., $D$. As a result, to satisfy~\eqref{power_contraint}, there exists a maximum value of the distance. By substituting $P_{\rm e}$ in~\eqref{power_contraint} with its calculation formula, the maximum distance can be calculated as the right-hand side of~\eqref{max_range}.


Consider $\Phi = \{\rho_{\rm c},X_{\rm c}^{\rm A}, X_{\rm s}^{\rm I}, X_{\rm s}^{\rm A},F^{\rm A}_{\rm e}, D^{\rm I,U}_{\rm max}\}$ as one of the optimal solutions if there exist more than one optimal solutions for Problem (P1). According to the proof in the first step, $D^{\rm I,U}_{\rm max}$ should be equal to or smaller than the term on the right-hand side of~\eqref{max_range} to make constraint~\eqref{SINR_constraint_in_problem} in Problem (P1) satisfied. As a result, we consider the following two cases: (i) In Case 1, $D^{\rm I,U}_{\rm max}$ is equal to the term on the right-hand side of~\eqref{max_range}. In this case, the considered optimal solution makes~\eqref{max_range} satisfied; (ii) In Case 2, $D^{\rm I,U}_{\rm max}$ is smaller than the term on the right-hand side of~\eqref{max_range}. We can thus increase $D^{\rm I,U}_{\rm max}$ to equal the term on the right-hand side of~\eqref{max_range}, which does not impact the overall resource consumption and ensures that all constraints in Problem (P1) remain satisfied.
Accordingly, the solution after the increase of $D^{\rm I,U}_{\rm max}$ is still optimal and makes~\eqref{max_range} satisfied.
If there exists one unique optimal solution for Problem (P1), for the optimal solution, $D^{\rm I,U}_{\rm max}$ must be equal to the term on the right-hand side of~\eqref{max_range}, i.e., making~\eqref{max_range} satisfied. This is because, otherwise, according to Case 2, we can find another optimal solution for (P1).

\section*{Appendix C\\ Proof of Proposition 3}
We respectively prove the three necessary conditions in~\eqref{necessary_opt_con}. First, defined in (\ref{upper_bound_NIU}), an upper bound of the expected number of targets that can be sensed by the mobile device is
\begin{equation}\label{E_N_s2I_appendix}
\overline{N}^{\rm U}_{\rm I}= \min\left\{\mathbb{P}\left\{D^{\rm I, U}\le D^{\rm I,U}_{\rm max}\right\}\cdot \mathbb{E}\left[N^{\rm U}_{\rm I,1}\right],\frac{1-\rho_{\rm c}}{\widehat{\rho}_{\rm s}}, \frac{\tau T\cdot F^{\rm I}_{\rm e}}{C_0} \right\},
\end{equation}
As a result, if $(1-\rho_{\rm c})/\widehat{\rho}_{\rm s}$ is greater than $\mathbb{E}[N^{\rm U}_{\rm I,1}]$ or $\tau T\cdot F^{\rm I}_{\rm e}/C_0$, we can find a greater $\rho_{\rm c}$ which decreases the amount of required spectrum resources, i.e., $X_{\rm c}^{\rm A}$, while keeping all the other parts of Problem (P1), including $\overline{N}^{\rm U}_{\rm I}$, unaffected. Therefore, for the optimal solutions to Problem (P1), $\rho_{\rm c}$ should satisfy the condition in (\ref{rho_c_new_constraint}):
\begin{equation}
\frac{1-\rho_{\rm c}}{\widehat{\rho}_{\rm s}} \le \min\left\{\mathbb{E}\left[N^{\rm U}_{\rm I,1}\right],\frac{\tau T\cdot F^{\rm I}_{\rm e}}{C_0}\right\}.
\end{equation}

Second, with the necessary condition, $(\ref{E_N_s2I_appendix})$ reduces to
\begin{equation}
\overline{N}^{\rm U}_{\rm I} = \min\left\{\mathbb{P}\left\{D^{\rm I, U}\le D^{\rm I,U}_{\rm max}\right\}\cdot \mathbb{E}\left[N^{\rm U}_{\rm I,1}\right],\frac{1-\rho_{\rm c}}{\widehat{\rho}_{\rm s}}\right\}.
\end{equation}
According to the expression derived in Section~\ref{subsec_decision_making}, $\mathbb{P}\left\{D^{\rm I, U}\le D^{\rm I,U}_{\rm max}\right\}$ increases with $D^{\rm I,U}_{\rm max}$ which increases with $X^{\rm I}_{\rm s}$ and decreases with $\rho_{\rm c}$, and $\mathbb{E}\left[N^{\rm U}_{\rm I,1}\right]$ is not affected by any decision variables. As a result, first, if $\mathbb{P}\left\{D^{\rm I, U}\le D^{\rm I,U}_{\rm max}\right\}\cdot \mathbb{E}\left[N^{\rm U}_{\rm I,1}\right]> (1-\rho_{\rm c})/\widehat{\rho}_{\rm s}$, we can find a smaller $X^{\rm I}_{\rm s}$ while keeping all the other parts of Problem (P1) including $\overline{N}^{\rm U}_{\rm I}$ unaffected. Second, if $\mathbb{P}\left\{D^{\rm I, U}\le D^{\rm I,U}_{\rm max}\right\}\cdot \mathbb{E}\left[N^{\rm U}_{\rm I,1}\right]< (1-\rho_{\rm c})/\widehat{\rho}_{\rm s}$, we can find a greater $\rho_{\rm c}$ which (i) increases $\overline{N}^{\rm U}_{\rm I}$ such as to decrease the amount of spectrum and computing resources required for target tracking by the APs; (ii) decreases the amount of required spectrum resources, i.e., $X_{\rm c}^{\rm A}$. To conclude, for the optimal solutions, the condition in (\ref{rho_c_XIs_new_constraint}) should be satisfied:
\begin{equation}
\frac{1-\rho_{\rm c}}{\widehat{\rho}_{\rm s}} = \mathbb{P}\left\{D^{\rm I, U}\le D^{\rm I,U}_{\rm max}\right\}\cdot \mathbb{E}\left[N^{\rm U}_{\rm I,1}\right].
\end{equation}

Third, for the necessary condition~\eqref{XsA_FeA_constraint}, if it is violated, we can find a smaller amount of required spectrum resources, i.e., $X_{\rm s}^{\rm A}$, or computing resources, i.e., $F_{\rm e}^{\rm A}$, without affecting all the other parts of Problem (P1).

\bibliographystyle{IEEEtran}
\bibliography{Hss_Ref}
\end{document}